\documentclass[12pt]{article}
\usepackage{amsmath}
\usepackage{graphicx}
\usepackage{enumerate}
\usepackage{natbib}
\usepackage{url} 

\newcommand{\blind}{0}

\usepackage{url} 
\usepackage{amsmath}
\usepackage{graphicx}
\usepackage{amsmath, amssymb}
\usepackage{bbold}
\usepackage[colorlinks=true,allcolors=blue]{hyperref}
\usepackage{mathbbol}
\usepackage{float}
\usepackage{xcolor}
\usepackage{bm} 
\usepackage{setspace}
\usepackage{multirow}
\usepackage{lscape}
\usepackage{rotating}
\usepackage{adjustbox}
\usepackage{booktabs}
\usepackage{ragged2e}
\usepackage[caption = false]{subfig} 
\usepackage{float}
\usepackage{amsthm}

\newcommand{\beginsupplement}{%
        \setcounter{equation}{0}
        \renewcommand{\theequation}{S\arabic{equation}}
         \renewcommand{\theHequation}{S\arabic{equation}}
         \setcounter{section}{0}
        \renewcommand{\thesection}{S\arabic{section}}
        \renewcommand{\theHsection}{S\arabic{section}}
        \setcounter{table}{0}
        \renewcommand{\thetable}{S\arabic{table}}%
        \renewcommand{\theHtable}{S\arabic{table}}
        \setcounter{figure}{0}
        \renewcommand{\thefigure}{S\arabic{figure}}
        \renewcommand{\theHfigure}{S\arabic{figure}}
     }

\newcommand{\bpm}{\begin{pmatrix}}
\newcommand{\epm}{\end{pmatrix}}

\theoremstyle{definition}
\newtheorem{definition}{Definition}[section]

\def\*#1{\mathbf{#1}}
\def\+#1{\amsmathbb{#1}}
\def\##1{\mathbb{#1}}
\DeclareSymbolFontAlphabet{\amsmathbb}{AMSb}%

\newcommand{\calT}{\mathcal T}

\newcommand{\bC}{ {\boldsymbol C} }

\newcommand{\bU}{ {\boldsymbol U} }

\newcommand{\bDelta}{ {\boldsymbol \Delta} }

\begin{document}

\def\spacingset#1{\renewcommand{\baselinestretch}%
{#1}\small\normalsize} \spacingset{1}

\if0\blind
{
  \title{\bf Functional Principal Component Analysis for Continuous non-Gaussian, Truncated, and Discrete Functional Data}
  \author{Debangan Dey
  \thanks{This study was supported by grant no. ZIA MH002954-07 Motor Activity Research Consortium for Health (mMARCH) from the National Institute of Health and a postdoctoral fellowship to Dr. Dey. The views and opinions expressed herein are those of the authors and should not be construed to represent the views of any of the sponsoring organizations, agencies, or the US government.}
    \hspace{.2cm}\\
    National Institute of Mental Health,\\
    Rahul Ghosal\\
    Department of Epidemiology and Biostatistics, University of South Carolina, \\
    Kathleen Merikangas\\
    National Institute of Mental Health, and \\
    Vadim Zipunnikov\\
    Department of Biostatistics, Johns Hopkins Bloomberg School of Public Health}
\maketitle
} \fi

\if1\blind
{
  \bigskip
  \bigskip
  \bigskip
  \begin{center}
    {\LARGE\bf Functional Principal Component Analysis for Continuous non-Gaussian, Truncated, and Discrete Functional Data}
\end{center}
  \medskip
} \fi

\bigskip

\begin{abstract}
\textcolor{black}{
Mobile health studies often collect multiple within-day self-reported assessments of participants' behavior and well-being on different scales such as physical activity (continuous), pain levels (truncated), mood states (ordinal), and life events (binary). These assessments, when indexed by time of day, can be treated as functional data of different types—continuous, truncated, ordinal, and binary. We develop a functional principal component analysis that deals with all four types of functional data in a unified manner. It employs a semiparametric Gaussian copula model, assuming a generalized latent non-paranormal process as the underlying mechanism for these four types of functional data. We specify latent temporal dependence using a covariance estimated through Kendall's $\tau$ bridging method, incorporating smoothness during the bridging process. Simulation studies demonstrate the method's competitive performance under both dense and sparse sampling conditions. We then apply this approach to data from 497 participants in the National Institute of Mental Health Family Study of the Mood Disorder Spectrum to characterize within-day temporal patterns of mood differences among individuals with major mood disorder subtypes, including Major Depressive Disorder, Type 1, and Type 2 Bipolar Disorder.
} 

\end{abstract}

\noindent%
{\it Keywords:}  Functional Data Analysis; EMA, Discrete Functional Data; Covariance Estimation; Gaussian copula
\vfill

\newpage
\spacingset{1.75} 

\section{Introduction}

\begin{figure}[ht]
\centering
\includegraphics[width=0.9\linewidth]{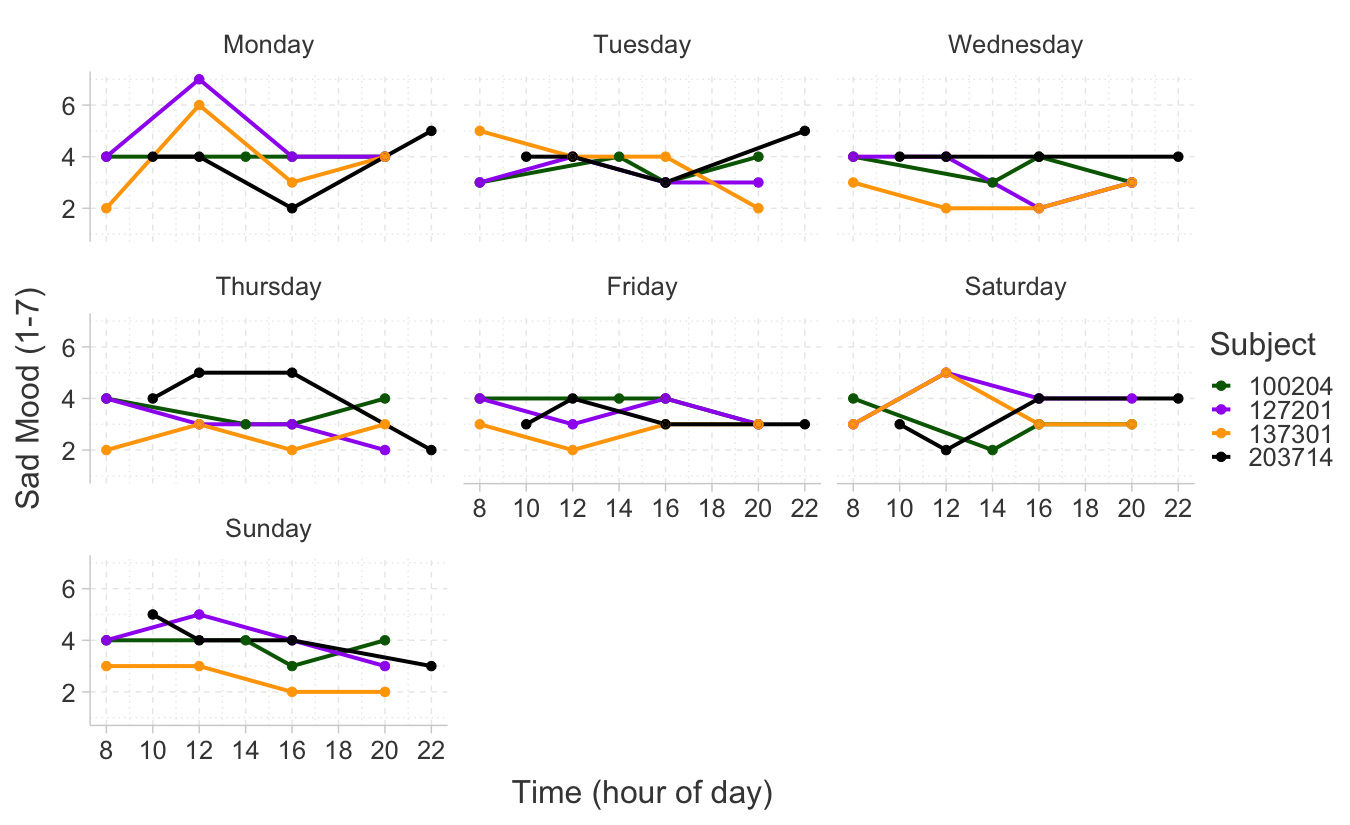}
\caption{Within-day temporal pattern of mood ratings ($1$ = Most Happy to $7$ = Most Sad, $4$ = Neutral) over $7$ days for four different subjects in National Institute of Mental Health Family Study of Mood Disorders Subtypes.}
\label{fig:motivation}
\end{figure}

Multiple real-time assessment of human physical and mental experiences and health-related behaviors has been recently made possible through mobile digital health monitoring (mHealth). Intensive longitudinal data collected with mHealth tools including wearables and smartphone apps allows researches to better understand within-day temporal patterns of  experiences and behaviours and their influence on health. National Institute of Mental Health Family Study of the Mood Disorder Spectrum employed an mHealth app that collected multiple within-day participants' reports on levels of sadness/happiness, energy, anxiety, pain, levels of physical activity, and many others experiences and behaviors. Self-reported mood was reported on a Likert ordinal scale of $(1)$ to $(7)$ (very happy to very sad, with (4) = neutral). Figure \ref{fig:motivation} displays diurnal (within-day) mood trajectories over four  time-points over seven days of the week for four randomly chosen participants.

\textcolor{black}{ The study also collected self-reports on binary (negative encounters: present/not present), truncated (pain: yes/no, and if yes, what is the level of pain on the scale 1-10), and multiple continuous scales covering quality and duration of sleep, time spent in physical activity of different intensity (light, moderate-to-vigorous). Indexed by time-of-day, these assessments can be treated as functional observations of continuous non-Gaussian, truncated, ordinal, and binary types. Figure \ref{fig:motivation} illustrates typical analytical challenges for these type of functional data: (i) they are sparsely observed, (ii) observation points can be highly irregular, (iii) there exists heterogeneity in subject-specific interpretations of Likert scales (specifically, for truncated, ordinal, and continuous scales). In this paper, we develop a covariance estimation and principal component analysis that can treat functional data of four types including continuous non-Gaussian, truncated, ordinal and binary in a unified way and addresses the above-mentioned statistical challenges. Next, we provide a review of relevant methods and place our work within existing literature.}



 The early developments in functional principal component analysis \citep{Ramsay05functionaldata, huang2004polynomial,yao2003shrinkage,yao2005functional,xiao2016fast,xiao2018fast} primarily focused on continuous Gaussian functional data collected under sparse or dense designs. These methods rely on smoothing noisy sample covariances \citep{besse1986principal,staniswalis1998nonparametric,yao2003shrinkage} or pre-smoothing the sample of functions and then diagonalizing the sample covariance function of the smoothed curves \citep{Ramsay05functionaldata,xiao2016fast}. 

\textcolor{black}{There are some interesting recent developments in fPCA for continuous non-Gaussian and non-continuous functional data. Proposed approaches mostly extend fPCA to continuous and non-continuous types generated from generalized exponential family via generalized linear models. The original work of \cite{hall2008modelling} relies on a known link function connecting observed generalized outcome to a latent continuous outcome and proceed with the covariance estimation of functional latent process}. These models have been extended further to accomodate conditional function-on-scalar regression models through frequentist \citep{gertheiss2015marginal, scheipl2016generalized, li2018exponential} and Bayesian paradigms \citep{van2009bayesian, goldsmith2015generalized}. \cite{li2014hierarchical} proposed to model multiple functional outcomes jointly and developed a hierarchical method for modeling paired functional data consisting of simultaneous measurements of a continuous and a binary variable. This approach modeled the correlation between the paired variables by the correlation across the principal component scores of latent ``pseudo'' normal distributions.  \textcolor{black}{All these approaches focus primarily on binary and count data, and rely either on known link functions or model-based likelihood}. \cite{wang2014generalized} developed an empirical Bayesian learning approach using a Gaussian approximation in the presence of ordinal outcomes. \cite{meyer2022ordinal} took a step further and used a probit link to model ordinal functional outcomes. Recently, \cite{leroux2023fast} developed a fast generalized functional principal component analysis that smooths local generalized linear models and requires dense functional data. 

 \textcolor{black}{In a seemingly related to our proposal, but conceptually different approach,  \cite{zhong2022robust} developed a robust functional principal component analysis method based on the Kendall’s $\tau$ function for continuous non-Gaussian functional data. Compared to our proposal, \cite{zhong2022robust} solely focuses on continuous functional data and uses a nonparametric kernel-based estimation method. \cite{solea2022copula} introduced a functional copula Gaussian graphical model for multivariate continuous functional data while assuming a nonparanormal distribution (Definition \ref{def:npn}) on the coefficients of the Karhunen–Loeve expansions.  
In this paper, we propose a unified functional principal component analysis for continuous non-Gaussian, truncated, ordinal, and binary functional data. The approach is rank-based and is, hence, invariant to arbitrary monotone transformations (and addresses subject-specific interpretation of reporting scales in mHealth applications) and  robust. Additionally, the approach computationally is very fast. Importantly, the method can accommodate both functional data collected under both dense and sparse designs.}

Our covariance estimation and principal component analysis approach is based on semiparametric Gaussian Copula. 
Beyond being applied to handle continuous and multivariate functional data \citep{ualbertaGaussianCopula,solea2022copula}, latent Semiparametric Gaussian Copula (SGC) \citep{liu2009Nonparanormal, liu2012high, fan2017high, yoon2018sparse, quan2018rank, dey2019connecting, huang2021latentcor,dey2022semiparametric} are powerful for modeling all four types (continuous, truncated, ordinal, and binary) of data and provide scale-free, robust and fast algorithms.
\textcolor{black}{
Here, we extend this framework to univariate functional data of continuous non-Gaussian, truncated, or discrete type.
Our key methodological contributions are: i) defining a data-generating mechanism for the observed continuous, truncated or discrete functional data via Generalized Latent Nonparanormal process, ii) developing a robust semiparametric method for estimating the smooth covariance of the latent Gaussian process based on the rank correlation (Kendall's $\tau$) of the observed data and directly incorporating smoothness within the bridging step; iii) extending the approach for handling both dense and sparse sampling designs, calculating subject-specific latent representations of observed data, latent principal components and latent principal component scores; iv) handling all four data types in a unified way. Importantly, in the case of continuous, truncated, and ordinal scales, the approach is scale-invariant.}

The rest of this article is organized as follows. Section 2 presents our modeling framework and illustrates the proposed covariance estimation method. In Section 3, we evaluate the performance of the proposed method via simulations and compare it with available existing methods of covariance estimation. Section 4 applies the method to mHealth data from NIMH Family Study. Section 5 presents concluding remarks and discusses possible extensions of this work.


\section{Modeling Framework}
We consider univariate functional data $X_i(t)$, for subjects $i=1,2,\ldots,n$, where $X_i(t)$ denotes the functional observation for subject $i$ and can be binary, ordinal, truncated or continuous. In this section, we assume that functional objects are observed on a dense and regular grid of points $S= \{t_{1},t_{2},\ldots,t_{m} \} \subset \mathcal{T}=[0,1]$. The proposed method is later extended to a more general setup of sparse and irregularly observed mixed-type functional data. Functional observations $X_{i}(\cdot)$  are assumed to be independent and identically distributed (i.i.d.) copies of $X(\cdot)$, an underlying smooth stochastic process. As in the case of continuous functional data, a major question of interest is estimating the covariance kernel of the process $X(\cdot)$. We propose using a Semiparametric Gaussian Copula model \citep{yoon2018sparse, dey2022semiparametric} for covariance modeling of continuous, truncated, or discrete type functional data. Next, we introduce the key components of the model.

\begin{definition}{(Nonparanormal distribution)}\label{def:npn} A random vector $X =(X_1,\dots,X_p)'$ follows a non-paranormal distribution denoted as $X\sim NPN_p(0,\Sigma,f)$, if there exists monotone transformation functions $f=(f_1,...,f_p)$ such that $L = f(X)=(f_1(X_1),\dots, f_p(X_p)) \sim N(0,\Sigma)$, with $\Sigma_{jj}=1$ for $1 \le j \le p$. The constraints on diagonal elements of $\Sigma$ are made to ensure the identifiability of the distribution as shown in \cite{liu2012high} and \cite{fan2017high}.
\end{definition}
\begin{definition}{(Generalized Latent Nonparanormal Process)}\label{def:glnpp}
For an observed process $X(t)$, we assume that there exists a latent process $Z(t)$ that serves as a data generating mechanism for observed $X(t)$ as follows: 
\begin{equation}
\begin{split}
 X(t) &= Z(t) \text{(continuous scale), or} \\
 X(t) &= Z(t) I(Z(t) > \Delta(t)), \text{(truncated scale), or}\\
 X(t) &= \sum_{k=0}^{l-1} k I(\Delta_k(t) \le Z(t) < \Delta_{k+1}(t)),  -\infty = \Delta_0(t) \le \Delta_1(t) \le \cdots \le \Delta_l(t)=\infty, \text{(ordinal scale), or}  \\
 X(t) &= I(Z(t) > \Delta(t)),\text{(binary scale)}.
\end{split}
\end{equation}

Suppose that for any finite set of points $S = \{t_{1},t_{2},\ldots,t_{m}\}, ((Z(t_1)),\cdots, (Z(t_m))) \sim NPN_{m}(0,\bC(S,S), f)$, where $f=(f_{t_1}, \cdots, f_{t_m})$ is a collection of $m$ monotone transformation functions and $(\bC(S,S))_{jj'} = C(t_j, t_{j'})$ for a scaled covariance function $C: \calT \times \calT \mapsto [-1,1]$ with $C(t,t)=1$. Further, we denote $V(t)$ as the latent multivariate normal process, i.e., $V(t) = f_t(Z(t))$ and $(V(t_1), \cdots, V(t_m)) \sim N(0, \bC(S,S))$. 
Then, 
$X(t)$ is defined as the Generalized Latent Nonparanormal Process (GLNPP) with latent covariance function $C$ and cutoff process $\bDelta(t)$. 
\end{definition}

\textcolor{black}{The definition \ref{def:glnpp} characterizes the GLNPP by describing the point-wise distribution of a finite collection of time points. However, this initial definition does not immediately reveal whether this point-wise characterization results in a stochastic process spanning an uncountable domain $\calT$. We can gain further insight by realizing that the finite-dimensional realization of $Z(t)$ is a Gaussian copula. Hence, we can express $f_t$ as $f_t = \Phi^{-1}(G_t(Z_t))$, where $G_t$ represents the pointwise cumulative distribution function (CDF) for the marginal distribution of $Z(t)$ and $\Phi$ denotes the standard Normal CDF \citep{liu2012high}. Drawing upon Corollary 5.11 from \cite{schmitz2003copulas}, we can ascertain that $Z(\cdot)$ constitutes a valid stochastic process across the entire domain. Given that $\bDelta(\cdot)$ is defined as a legitimate stochastic process, we can consequently affirm that $X(\cdot)$ also qualifies as a valid process spanning the entirety of the domain.}

For finite-dimensional multivariate distribution, the efficient way to estimate latent correlation $C(t_j,t_{j'})$ is to use Kendall's $\tau$ rank correlation \citep{fan2017high,yoon2018sparse, quan2018rank, huang2021latentcor}. The population Kendall's $\tau$ between observations at $j$-th and $j'$-th time-point can be defined as $\tau_{jj'}= E(sgn\{(X_i(t_j)-X_{i'}(t_{j}))(X_i(t_{j'}) - X_{i'}(t_{j'}))\})$, where $X_i(\cdot)$ and $X_{i'}(\cdot)$ are two independent realizations of the process. The population Kendall's $\tau$ has shown to be related to the population latent correlation parameter through a bridging function $F(\cdot)$ both for pairs of the same type and their combinations in \citep{yoon2018sparse, dey2022semiparametric}, i.e., $\tau_{jj'} = F(C(t_j, t_{j'}))$. The bridging function $F$ depends on the type of variable pair and the marginal cutoffs. 
\subsection{Covariance Estimation}
We model the scaled covariance function $C$ (note that $C(t,t)=1$ by definition) of the latent process as a tensor product spline on $\calT \times \calT$ - 
\begin{equation}
C(s,t) = g(\sum_{k=1}^{d}\sum_{l=1}^{d} u_{kl} B_k(s) B_l(t)), s\ne t,
\label{eq: cst}
\end{equation}
Here, ${B_1(\cdot), . . . , B_d(\cdot)}$ is a collection of basis functions on $\calT$. 
In this article, we use cubic B-spline basis functions; other basis functions can also be used depending on the problem of interest. Here $\bU= (u_{kl})_{1\leq k \leq d, 1\leq l \leq d}$ denotes the unknown coefficient matrix and $g = \frac{e^x-1}{e^x+1}$ is an inverse Fischer transformation to ensure the correlation values are within $[-1,1]$. We also enforce the constraint $u_{kl}=u_{lk}$ on the coefficient matrix $\bU$ to ensure $C(s,t)=C(t,s)$. Ideally, if we had observed the latent continuous process $V(\cdot)$, we would like to find $\bU$ that minimizes the following nonlinear least square objective function  - 

\begin{equation}
l= \sum_{i=1}^{n} \sum_{1\leq j < j' \leq m} (V_i(t_j) V_i (t_{j'}) - C(t_{j},t_{j'}))^2 = \sum_{i=1}^{n} \sum_{1\leq j < j' \leq m} (V_i(t_j) V_i (t_{j'}) - g(\sum_{k=1}^{d}\sum_{l=1}^{d} u_{kl} B_k(t_j) B_l(t_{j'})))^2
\label{eq: obj-latent}
\end{equation}

Instead, we can translate the objective function in \eqref{eq: obj-latent} in terms of observed process $X(\cdot)$ and the Kendall's $\tau$ correlation since Kendall's $\tau$ is invariant to monotonic transformations. First, let's denote $\delta_{jj'}^{ii'} = sgn\{(X_i(t_j)-X_{i'}(t_{j}))(X_i(t_{j'}) - X_{i'}(t_{j'}))\})$ and note that, $E(\delta_{jj'}^{ii'})= \tau_{jj'}$. We also define the sample Kendall's $\tau$ as: $\hat{\tau}_{jj'} = \frac{2}{n(n-1)} \sum_{1\le i < i' < n} sgn\{(X_i(t_j)-X_{i'}(t_j))(X_i(t_{j'}) - X_{i'}(t_{j'}))\} = \frac{2}{n(n-1)} \sum_{1\le i < i' < n} \delta_{jj'}^{ii'} = \bar{\delta}_{jj'}$. Let us also denote $\delta_{jj'}= F(C(t_j, t_{j'}))= F(g(\sum_{k=1}^{d}\sum_{l=1}^{d} u_{kl} B_k(t_j) B_l(t_{j'})))$. We seek to find $\bU$ that minimize 
\begin{equation}
\begin{split}
\tilde{l}& = \sum_{1 \leq i < i' <n} \sum_{1\leq j < j' \leq m} (\delta_{jj'}^{ii'} - \delta_{jj'})^2\\
& = \sum_{1 \leq i < i' <n} \sum_{1\leq j < j' \leq m} (\delta_{jj'}^{ii'} - \bar{\delta}_{jj'} + \bar{\delta}_{jj'} - \delta_{jj'})^2\\
& = \sum_{1 \leq i < i' <n} \sum_{1\leq j < j' \leq m} (\delta_{jj'}^{ii'} - \bar{\delta}_{jj'})^2 + \sum_{1\leq j < j' \leq m} {n\choose2} (\bar{\delta}_{jj'} - \delta_{jj'})^2\\
& = \sum_{1 \leq i < i' <n} \sum_{1\leq j < j' \leq m} (\delta_{jj'}^{ii'} - \bar{\delta}_{jj'})^2 + {n\choose2} \sum_{1\leq j < j' \leq m}  (\hat{\tau}_{jj'} - F(g(\sum_{k=1}^{d}\sum_{l=1}^{d} u_{kl} B_k(t_j) B_l(t_{j'}))))^2,
\end{split}
\label{eq: obj-obs}
\end{equation}
where $\tilde{l}=\tilde{l}(\bU)$ and $\delta_{jj'}= F(C(t_j, t_{j'}))= F(g(\sum_{k=1}^{d}\sum_{l=1}^{d} u_{kl} B_k(t_j) B_l(t_{j'})))$. Only the second term of the objective function $\tilde{l}$ in \eqref{eq: obj-obs} depends on $\bU$. Hence, minimizing $\tilde{l}$ is equivalent to minimizing the nonlinear least square objective function $\sum_{1\leq j < j' \leq m}  (\hat{\tau}_{jj'} - F(g(\sum_{k=1}^{d}\sum_{l=1}^{d} u_{kl} B_k(t_j) B_l(t_{j'}))))^2$ with respect to $\bU$. This minimization problem hence  translates to fitting a regression model using non-linear least squares \citep{bates1988nonlinear}
to obtain estimate of the coefficients $u_{kl}$. 
In particular, we have,
\begin{equation}\label{eq:obj-tau}
      \hat{\bU}= \underset{
        \bU}{argmin}  \hspace{2 mm} \sum_{1\leq j < j' \leq m}  (\hat{\tau}_{jj'} - F(g(\sum_{k=1}^{d}\sum_{l=1}^{d} u_{kl} B_k(t_j) B_l(t_{j'}))))^2
\end{equation}
Subsequently, $\hat{C}(s,t) = g(\sum_{k=1}^{d}\sum_{l=1}^{d} \hat{u}_{kl}B_k(s) B_l(t)), s\ne t,$  is the estimator of the correlation function ($\hat{C}(t,t)=1$). The Gauss–Newton algorithm within \texttt{nls} function in R is used to solve the non-linear optimization problem. The choice of the number of basis functions $d$ controls the smoothness of the estimated latent correlation. In this article, we follow a truncated basis approach \citep{Ramsay05functionaldata,fan2015functional}, by restricting the number of B-spline bases to be small in both directions to incorporate smoothness. The development of a regularized estimation approach of the latent correlation with smoothness constraint will be pursued in future research.

\textcolor{black}{\subsection*{Bridging Functions}}
\textcolor{black}{In our estimation process, we need to know the bridging function introduced in Equation \eqref{eq:obj-tau}. Analytic expressions for the bridging functions are available in the literature for various scenarios, including continuous and binary variables \citep{liu2012high}, truncated variables \citep{yoon2018sparse}, and ordinal variables \citep{dey2022semiparametric}. Given our focus on the marginal distribution of univariate mixed processes (continuous/truncated/ordinal/binary), it is imperative to ascertain the specific analytic forms of the bridging functions that facilitate the transformation of Kendall's $\tau$ to latent correlation for pairs of variables falling into the same category (e.g., continuous-continuous, truncated-truncated). We provide these analytic expressions in Appendix A of the Supplementary Material.
}

\subsection*{Estimation of the Cutoff Parameters}
Note that the bridging function $F(\cdot)$ depends on the cutoffs for pairs involving binary, ordinal, and truncated functional data \citep{dey2022semiparametric,fan2017high}. Hence, we follow a method of moments-based estimation approach described in \cite {dey2022semiparametric} to get point-wise estimates of the cutoff process $\bm\Delta(\cdot)$. Observe that, the cutoff process is only identifiable up to a monotone transformation \citep{liu2012high}. Hence, we can only get estimates of $f_t(\bDelta(t))$ but we do a slight abuse of notation to denote the cutoffs as $\bDelta(t)$ in the estimating equations. From the observed data, we can estimate the cutoffs through the method of moments as follows:

\begin{equation}
\begin{split}
\textrm{Binary: } \widehat{\Delta}(t) & =\Phi^{-1}\left(\frac{\sum_{i=1}^{n}I(X_{i}(t)=0)}{n}\right)\\
\textrm{Ordinal: } \widehat{\Delta}_k(t) & = \Phi^{-1}\left(\frac{\sum_{i=1}^{n}I(X_{i}(t)<=(k-1))}{n}\right), k=1, \dots, l-1\\
\textrm{Truncated: } \widehat{\Delta}(t) & =\Phi^{-1}\left(\frac{\sum_{i=1}^{n}I(X_{i}(t)=0)}{n}\right)
\end{split}
 \label{eqn:cutoff-est}
\end{equation}

\textcolor{black}{\subsection*{Estimation of the transformation functions $f_t$}}
We don't need to know the monotone transformation functions for estimating the covariance parameters, and the transformation functions aren't estimable for binary and ordinal values \citep{dey2022semiparametric, liu2012high}. However, we need to get a point-wise estimate of the transformation function to get latent predictions at the time points for continuous and truncated variables (for non-zero observed value). We define -   

\begin{equation}
\begin{aligned}\label{ecdfct}
\hat{G}_{t}(x) = \frac{1}{n+1} \sum_{i=1}^{n} I(X_{i}(t) \le x), x \in \mathbb{R}, \text{(for continuous process)} \\
\hat{G}_{t}(x) = \frac{1}{n+1} \sum_{i=1}^{n} I(X_{i}(t) \le x), x > 0, \text{(for truncated process)}\\
\end{aligned}
\end{equation}
Then, we can use equation \eqref{ecdfct} to get the estimator of monotone transformations as follows (based on Section 4 of \cite{liu2009Nonparanormal}) -- 
\begin{equation}
\begin{aligned}\label{fct}
\hat{f}_t(x) = \Phi^{-1}(\hat{G}_{t}(x)), \text{(for continuous process)} \\
\hat{f}_t(x) = \Phi^{-1}(\hat{G}_{t}(x)),\text{(for truncated process)}\\
\end{aligned}
\end{equation}

\subsection{Extension to Sparse Data}
The proposed estimation method could be extended to more general scenarios where each curve $X_i(\cdot)$ can be observed sparsely and irregularly at time points $S_i=\{t_{i1},t_{i2},\ldots, t_{im_i}\}$. Although,
individual number of observations $m_{i}$ is small, we consider the scenario where $\bigcup_{i=1}^{n}\bigcup_{j=1}^{m_i}{t_{ij}}$ is dense in  $[0,1]$ \citep{kim2016general}. Let us denote $\bigcup_{i=1}^{n}\bigcup_{j=1}^{m_i}{t_{ij}}= \{t_1,t_2,\ldots,t_M\}$. In this case we can get the sample Kendall's $\tau$ as: $\hat{\tau}_{jj'} =\bar{\delta}_{jj'}= \frac{1}{N_{jj'}} \sum_{1\le i < i' < n: n^{ii"}_{jj'}> 0} sgn\{(X_i(t_j)-X_{i'}(t_j))(X_i(t_{j'}) - X_i(t_{j'}))\} = \frac{1}{N_{jj'}} \sum_{1\le i < i' < n: n^{ii"}_{jj'}> 0} \delta_{jj'}^{ii'} $. Here $n^{ii"}_{jj'}=1$ if both $X_i(\cdot),X_{i'}(\cdot)$ are observed at time-points $t_j$ and $t_{j'}$ and $N_{jj'}$ denote the total number of pairwise complete observation at $t_j,t_{j'}$. Based on the least square objective function from (4), we define the modified objective function for sparse data as,
\begin{equation}
\begin{split}
\tilde{l}(\bU)& = \sum_{1\leq j < j' \leq M : N_{jj'} >c_0 }  (\hat{\tau}_{jj'} - F(g(\sum_{k=1}^{d}\sum_{l=1}^{d} u_{kl} B_k(t_j) B_l(t_{j'}))))^2
\end{split}
\label{eq: sparse obj-obs}
\end{equation}
Here, we only consider those time points for which there are a sufficient number of pairwise observations, $c_0$, to estimate $\tau_{jj'}$ accurately. Throughout the article, we set $c_0=5$, which has resulted in a satisfactory performance, as demonstrated in our empirical analysis.

\vspace*{- 4mm}
\subsection{Curve Prediction}
Suppose the curve $X_i(\cdot)$ is observed at time-points $\{t_{i1},t_{i2},\ldots, t_{im_i}\}$ and we want to predict
$X_i(\cdot)$ at $\{s_{i1},s_{i2},\ldots, s_{im}\}$. This could be important in predicting the trajectory of a subject with partial information. Let us denote the latent observations corresponding to observed data by $\*V_i^O$ and new data by $\*V_i^N$. On latent scale, $\begin{pmatrix}
\*V_i^O\\
\*V_i^N
\end{pmatrix}\sim N(\begin{pmatrix}
\*0\\
\*0
\end{pmatrix},\begin{pmatrix}
\+C^{O,O} & \+C^{O,N}  \\
\+C^{N,O} & \+C^{N,N} 
\end{pmatrix})$. The individual components of the covariance matrix are obtained from the estimated covariance matrix using methods from Section 2.1. We have 
$$E(\*V_i^N\mid \*V_i^O )= \+C^{N,O}\{\+C^{O,O}\}{^{-1}} (\*V_i^O-\*0 ).$$
$$Cov(\*V_i^N\mid \*V_i^O )=\+C^{N,N} - \+C^{N,O}\{\+C^{O,O}\}{^{-1}} \+C^{O,N}.$$

Latent predictions can be obtained as  $\hat{\*V_i}^N=\hat{\+C}^{N,O}\hat{\+C}^{O,O}{^{-1}} (\hat{\*V_i}^O-\*0 )$, where $\hat{\*V_i}^O=E(V_i^0(\cdot)|X_i^0(\cdot))$ is the best linear unbiased predictor (BLUP) based on the proposal of \cite{dey2022semiparametric}.
$\hat{\*X_i}^N$ can be obtained from $\hat{\*V_i}^N$ using the estimated cutoff parameters and the estimated transformation functions. For example for binary $X(\cdot)$, we would have $\hat{\*X_i}^N(t)=I(\hat{\*V_i}^N(t)>\hat{\Delta}_t)$. 

\subsection{Latent Principal Components and Scores}
\textcolor{black}{
For continuous functional data $X_i(t)$, we can use truncated Karhunen-Lo$\grave{e}$ve expansion of $X_{i}(\cdot)$ as
${X}_{i}(t) \approx {\mu}(t)+\sum_{k=1}^{K}{\zeta}_{ik}{
\psi}_{k}(t)$, where ${\zeta}_{ik}$ are mean zero functional principal component (fPC) scores with variance $\lambda_k$ and $\psi_{k}(t)$ are orthogonal eigenfunctions \citep{yao2005functional}. Let $\Sigma(s,t)$ denote the covariance function of $X(\cdot)$, i.e., $\Sigma(s,t)=Cov(X(s),X(t))$. By Mercer's theorem, the covariance kernel has the following spectral decomposition $\Sigma(s,t)=\sum_{k=1}^{\infty}\lambda_k\psi_k(s)\psi_k(t)$.
The eigenvalues and eigenfunctions hence can be estimated from the eigen-equation $\int_{0}^{1}\Sigma(s,t)\phi_k(t)dt=\lambda_k\psi_k(t)$. This procedure is referred to as functional principal component analysis (FPCA). 
Uncorrelated scores ${\zeta}_{ik}$ serve as a multivariate summary of infinite dimensional data and are widely used in many applications.} 

\textcolor{black}{
For truncated or discrete functional data $X(t)$, the KL expansion is not directly applicable. We can instead use the estimated latent correlation function $\hat{C}(s,t)$ to perform FPCA. We use the eigenequation $\int_{0}^{1}\hat{C}(s,t)\psi_k^L(t)dt=\lambda_k\psi_k^L(t)$, to estimate the latent eigenfunctions, which reveals the major modes of variations in the data. The scores of the latent representations $V(t)$ can be obtained using standard FPCA techniques \citep{yao2005functional}, if $V(t)$ was available. To obtain latent functional prinicipal component (fPC) scores, we use the estimated smoothed covariance function $\hat{C}(s,t)$ to obtain BLUPs of the latent representations $\hat{V}(\cdot)=E(V(\cdot)|X(\cdot))$ \citep{dey2022semiparametric} and use FPCA on the latent representations $\hat{V}(t)$ to estimate the latent fPC scores ${\zeta}_{ik}^L$.}


\vspace*{- 4mm}
\subsection{Distances between mixed-type functional data}
Distances between two mixed-type functional curves (of the same type) $X_i(\cdot), X_j(\cdot)$ can be defined based on the distances between latent trajectories  $V_i(\cdot), V_j(\cdot)$, which is same as the Euclidean distance between estimated latent scores $\{{\zeta}_{ik}^L,{\zeta}_{jk}^L\}_{k=1}^K$ as, 
\begin{equation}
    d(X_i(\cdot),X_j(\cdot))=||\bm\zeta_i^L-\bm\zeta_j^L||_2.
\end{equation}
This metric captures the distance between latent trajectories $V_i(\cdot), V_j(\cdot)$ due to the orthogonality of the latent eigenfunctions.
This provides a novel way to extend distance-based clustering approaches \citep{jacques2014functional} to mixed-type functional data. The latent trajectories and their distances could be further explored to reveal various group-level patterns.

\section{Simulation Study}
\label{simul}
In this Section, we investigate the performance of the proposed estimation method via numerical simulations. To this end, the following scenarios are considered.

\subsection{Data Generating Scenarios}
In simulations, we consider four data types -  binary, ordinal, truncated, and continuous. For each data type, the latent functional process $V(\cdot)$ is generated from either of the following two correlation functions. 

\subsection*{Stationary Covariance $C_1(\cdot,\cdot)$}
In this case $V(\cdot)\sim \mathcal{GP}(0,C_1(\cdot,\cdot))$, where \textcolor{black}{$\mathcal{GP}(\mu(\cdot), C(\cdot,\cdot))$ denotes a Gaussian process with mean function $\mu(\cdot)$ and covariance function $C(\cdot, \cdot)$}. Here, we take $C_1(s,t)$ to be the Matern correlation kernel with parameters $\sigma^2=1$, $\nu=3.5$ and $\tau=\frac{1}{7}$. 
The covariance kernel of a general Mat\'ern process with parameters $\nu,\sigma^2,\tau$ is given by $C_1(s,t)=\sigma^2 \frac{2^{1-\nu}}{\Gamma(\nu)}(\sqrt{2\nu}\frac{d}{\tau})^\nu K_{\nu}(\sqrt{2\nu}\frac{d}{\tau}),  \hspace{2mm} d=|s-t|$, where $K_{\nu}$ is the modified Bessel function of the second kind.

\subsection*{Non-stationary Covariance $C_2(\cdot,\cdot)$}
In this case, the latent process $V(t)$ is again generated as $V(\cdot)\sim \mathcal{GP}(0,C_2(\cdot,\cdot))$, where $C_2(\cdot,\cdot)$ 
is a non-stationary correlation function given by $C_2(s,t)=\frac{C_2^*(s,t)}{\sqrt{C_2^*(s,s)}\sqrt{C_2^*(t,t)}}$ and the covariance function $C_2^*(s,t)=0.05^2 2sin(\pi t) sin(\pi s)+0.09^2 2cos(\pi t) cos(\pi s)+0.01^2I(s=t)$.
\subsection*{Scenario A: Binary Functional Data}
In this case, the observed functional data $X(t)$ is generated from the latent functional data $V(t)$ as $X(t)=I(V(t)>\Delta_t)$, where $\Delta_t=2.5$ is used for all $t$. We consider a dense and equispaced grid of $m=50$ time-points $\mathcal{S}$ in $\mathcal{T}=[0,1]$ for the observed data $X(t)$. Two sets of correlation function $C_1(\cdot,\cdot)$, and three sets of sample size $n\in \{100,500,1000\}$ are considered for data generation. We also consider an additional simulation scenario with $n=1000$, where each curve $X_i(t)$ is observed sparsely on a set of randomly chosen $10\%$ time-points from $\mathcal{S}$. For the stationary covariance function, particularly for this sparse scenario, we use a Matern correlation kernel with parameters $\sigma^2=1$, $\nu=3.5$ and $\tau=\frac{1}{2}$. The rest of the design is kept exactly the same. 

\vspace*{- 6mm}
\subsection*{Scenario B: Ordinal Functional Data}
In this case, the observed functional data $X(t)$ is generated as, 
\begin{equation*}
X(t)=
\begin{cases}
0\hspace{1.4 cm} \text{if $-\infty\leq V(t)< -0.6$}.\\
1\hspace{1.4 cm} \text{if $-0.6\leq V(t)< 0.1$}.\\
2\hspace{1.4 cm} \text{if $0.1\leq V(t)< 0.6$}.\\
3\hspace{1.4 cm} \text{if $0.6\leq V(t)< \infty$}.\\
\end{cases}
\end{equation*}
The rest of the simulation design (dense grid), including covariance kernels and sample sizes, are kept exactly the same as in scenario A.

\subsection*{Scenario C: Truncated Functional Data}
The observed functional data $X(t)$ is generated based on the latent process $V(\cdot)$ as,
\begin{equation*}
X(t)=
\begin{cases}
0\hspace{2 cm} \text{if $V(t)< 0.5$}.\\
$V(t)$\hspace{1.4 cm} \text{if $0.5\leq V(t)$}.\\
\end{cases}
\end{equation*}
The rest of the simulation design is again kept the same as in scenario A (dense).

\subsection*{Scenario D: Continuous Functional Data}
The observed functional data $X(t)$ is generated as a continuous transformation of latent process $V(\cdot)$ as,
\begin{equation*}
X(t)= \{V(t)\}^3.
\end{equation*}
The rest of the simulation design is kept the same as in scenario A (dense).

\subsection{Simulation Results}
We generated 100 Monte-Carlo (M.C) replications from each of the simulation scenarios to assess the performance of the proposed estimation method. 

\subsection*{Performance under scenario A:}
We use 7 cubic B-spline basis functions (over both arguments) to model the latent correlation function $C(s,t)$ using tensor product splines on $\calT \times \calT$. We compare our proposed approach (denoted henceforth as FSGC) to a few alternatives: i) naive FPCA using observed curves $X(t)$, \textcolor{black}{which captures the within-curve correlation in the observed data} ii) binary FPCA (bfpca) proposed by \cite{wrobel2019registration}. We also use FPCA on the latent predictions from multivariate (non-functional) SGCRM \citep{dey2022semiparametric} (referred henceforth as FSGC Latent). \textcolor{black}{Note that FSGC Latent is essentially a naive version of our proposed method treating functional observations as multivariate and smoothing performed in the last step. This could be inefficient, for example in sparse and irregular designs.} 
We compare the performance of FSGC with bfpca and FSGC Latent in terms of the integrated square errors for estimation of the covariance functions, where ISE is defined as $ISE=\int_0^1\int_0^1\{C_T(s,t)-\hat{C}(s,t)\}^2dsdt$, where $C_T(s,t)$ is the true covariance function. \textcolor{black}{The results from naive FPCA is not numerically comparable with FSGC as it captures the correlation structure in the observed data, hence we only report the estimated covariance surfaces which gives us an idea of what the estimated covariance surface would be, if we treated the observed data as continuous functional data.}

\begin{figure}[H]
\centering
\includegraphics[width=.9\linewidth , height=1\linewidth]{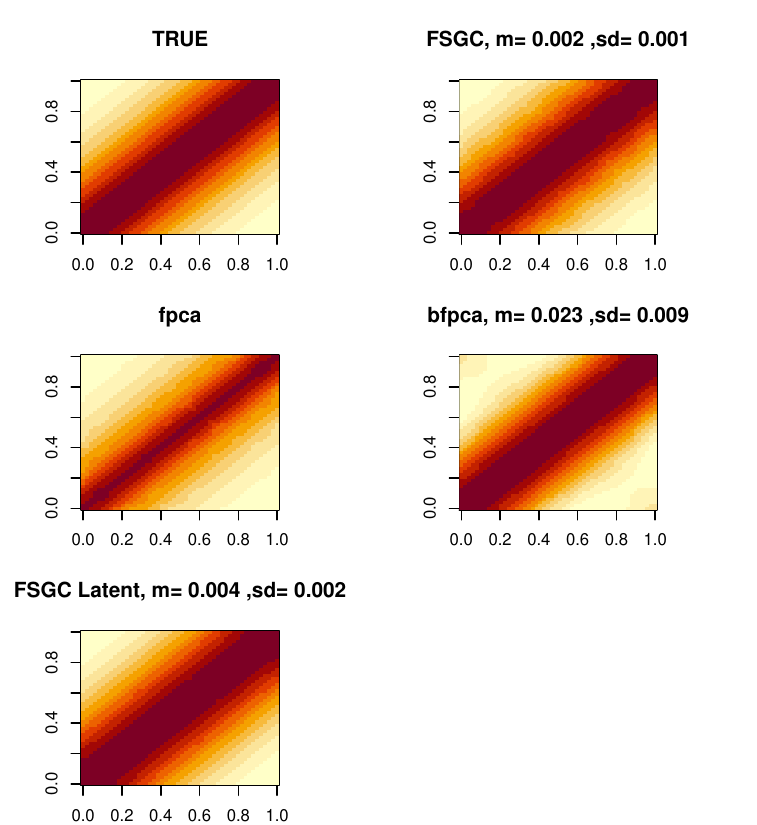}
\caption{True and estimated average covariance surface for stationary covariance kernel, scenario A, n=500. Average ISE (and sd) of the estimates are reported on the top of each image. FSGC denotes the proposed FSGC method, fpca denotes naive FPCA of the observed data, bfpca denotes binary FPCA, FSGC Latent denotes FPCA on latent predictions from SGCRM.}
\label{fig:fig1}
\end{figure}

Figure \ref{fig:fig1} displays the average estimated covariance surface over the grid $\calT \times \calT$ from the four estimation approaches along with the true covariance surface for the stationary covariance case and $n=500$. The mean (m) and standard deviation (sd) of ISE are reported for all cases using MC replications. We observe that the proposed FSGC method clearly outperforms binary FPCA in terms of estimation accuracy based on the average ISE. In particular, the proposed FSGC approach is seen to produce 10 times smaller ISE. The average ISE from the FSGC approach is also found to be 2 times smaller than the FSGC Latent approach, highlighting the importance of smoothing. 

Figure \ref{fig:fig2} displays the average estimated covariance surface and the estimates for the non-stationary covariance case and $n=500$. We observe that the proposed FSGC method produces superior performance compared to the binary FPCA method. The average ISE from the direct FSGC approach is found to be comparable to the FSGC Latent approach in this case. The results for $n=100,1000$ are similar for the stationary and non-stationary scenarios, where a superior performance of the FSGC method can be noted. The results are reported in Figures S1, S2, S4 and S5 of the Supplementary Material. 

We also apply the proposed FSGC method to estimate the latent scores as outlined in Section 2.4. The FSGC latent (multivariate) approach is also used. Note that the latent curves are identifiable up to a monotone transformation only. In Supplementary Table S1, we report the correlation between the estimated latent scores and the estimated scores (using FPCA) from the true latent curves. We observe that the first two scores are highly correlated with their estimated counterparts across all sample sizes and covariance types.
The results from the additional simulation scenario (sparse design) along with the performance of the curve prediction method are reported in Appendix B of Supplementary Material, where a superior performance of the FSGC can be noticed. 


\subsection*{Performance under scenario B:}
We apply the proposed FSGC method along with i) FPCA on the observed curve $X(t)$ and ii) FSGC Latent method to estimate the latent correlation function $C(s,t)$. The performance of the estimators is reported in terms of the integrated square errors (except naive FPCA). Supplementary Figure S8-S13 display the average estimated correlation surface over the grid $\calT \times \calT$ from the three estimation approaches along with the true surface for the stationary and non-stationary covariance kernels and sample sizes $n=100,500,1000$ respectively. The mean (m) and standard deviation (sd) of the ISE are reported using all the M.C replications. It can be observed that for this ordinal scenario, both FSGC and FSGC Latent perform comparably for stationary and non-stationary covariance. The performance of the estimators in terms of ISE improves with increasing sample sizes.

\subsection*{Performance under scenario C,D:}
The qualitative and quantitative performance of the proposed estimation method under scenarios C, and D are similar and are reported in Figures S14-S25 of the Supplementary Material. 

Overall, the proposed FSGC method can be observed to estimate the true covariance function accurately across all data types and covariance scenarios and thus provides a unified framework for covariance estimation for continuous, truncated, and discrete functional data.

\section{Real Data Application}
We apply our approach to mHealth data collected in the National Institute of Mental Health Family Study of the Mood Disorder Spectrum \citep{merikangas2014independence,stapp2022comorbidity}. The study is a large community-based study with participants recruited from a community screening of the greater Washington, DC metropolitan area. 
Our analysis focuses on a sample of 497 participants with ages ranging from 7 to 84 years old and other demographics reported in Supplementary Table S1.

Participants used a smartphone app to rate their current mood on a Likert scale from (1) to (7) (very happy to very sad, with (4) = neutral) four times per day during 7 AM - 11 PM time period each day for fourteen consecutive days. Ratings (5),(6), and (7) were collapsed into a single group for this analysis due to a very small number of participants reporting in those ranges, resulting in $5$ ordinal categories for emotional states. Our key research interest is to understand better within-day temporal patterns of reported mood states and whether differences in these patterns are associated with affective disorders. 
For our analysis, we focus on the midpoints of sixteen one-hour windows starting at 7 AM and ending at 11 PM (the windows are 7 AM -- 8 AM, 8 AM -- 9 AM, $\cdots$, 10 PM -- 11 PM with corresponding midpoints at 7:30 AM, 8:30 AM, etc). We consider $6591$ subject-day-level functional data of ordinal type across all the subjects. Note that subject-level clustering
is ignored and will be pursued in future as an extension of the current approach to multilevel functional data. In principle, this resembles a marginal FPCA approach \citep{park2015longitudinal} but is applied to ordinal functional data using our proposed method.

\begin{figure}[H]
\begin{center}
\begin{tabular}{ll}
\includegraphics[width=0.8\linewidth , height=0.3\linewidth]{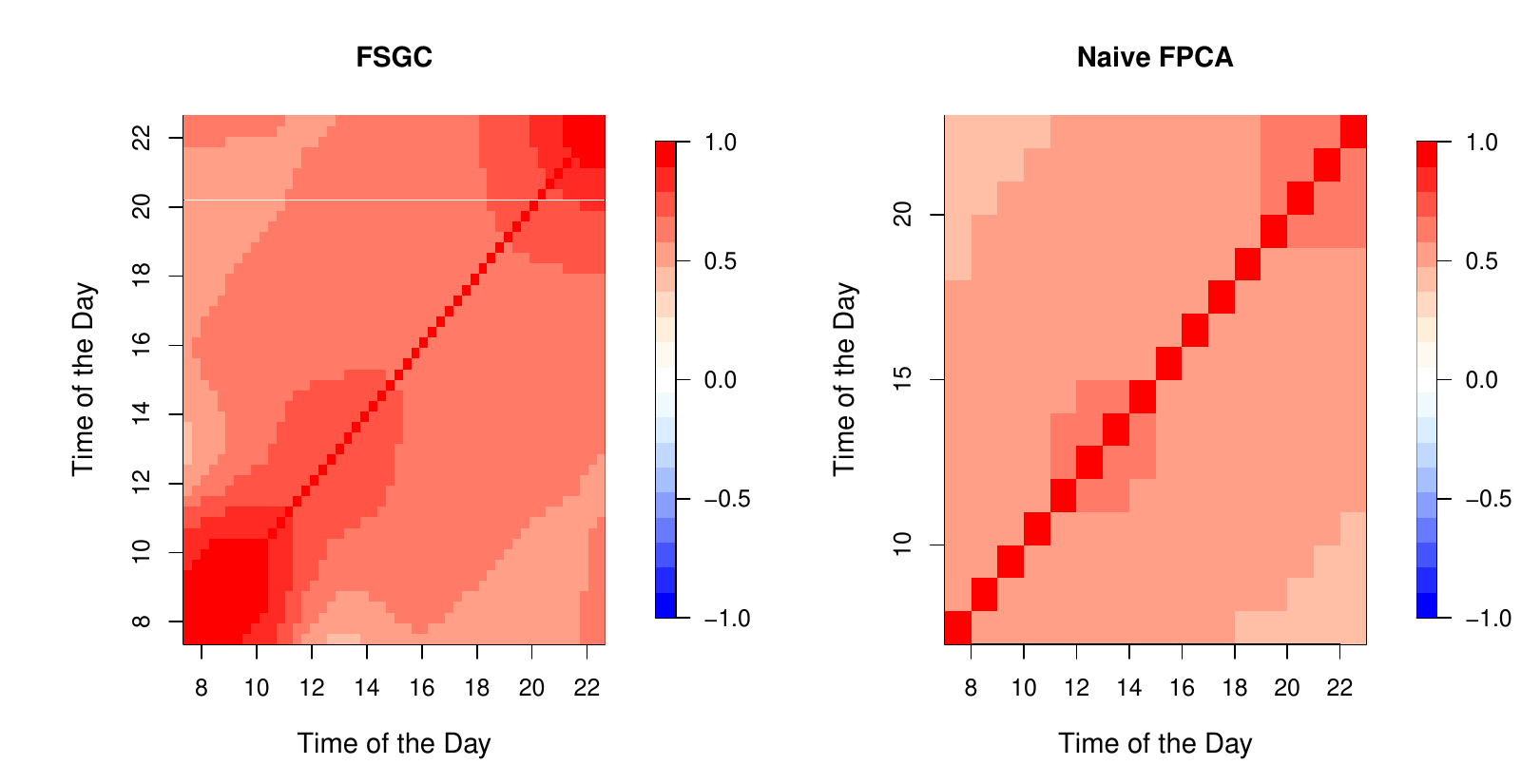} \\
\includegraphics[width=0.8\linewidth , height=0.55\linewidth]{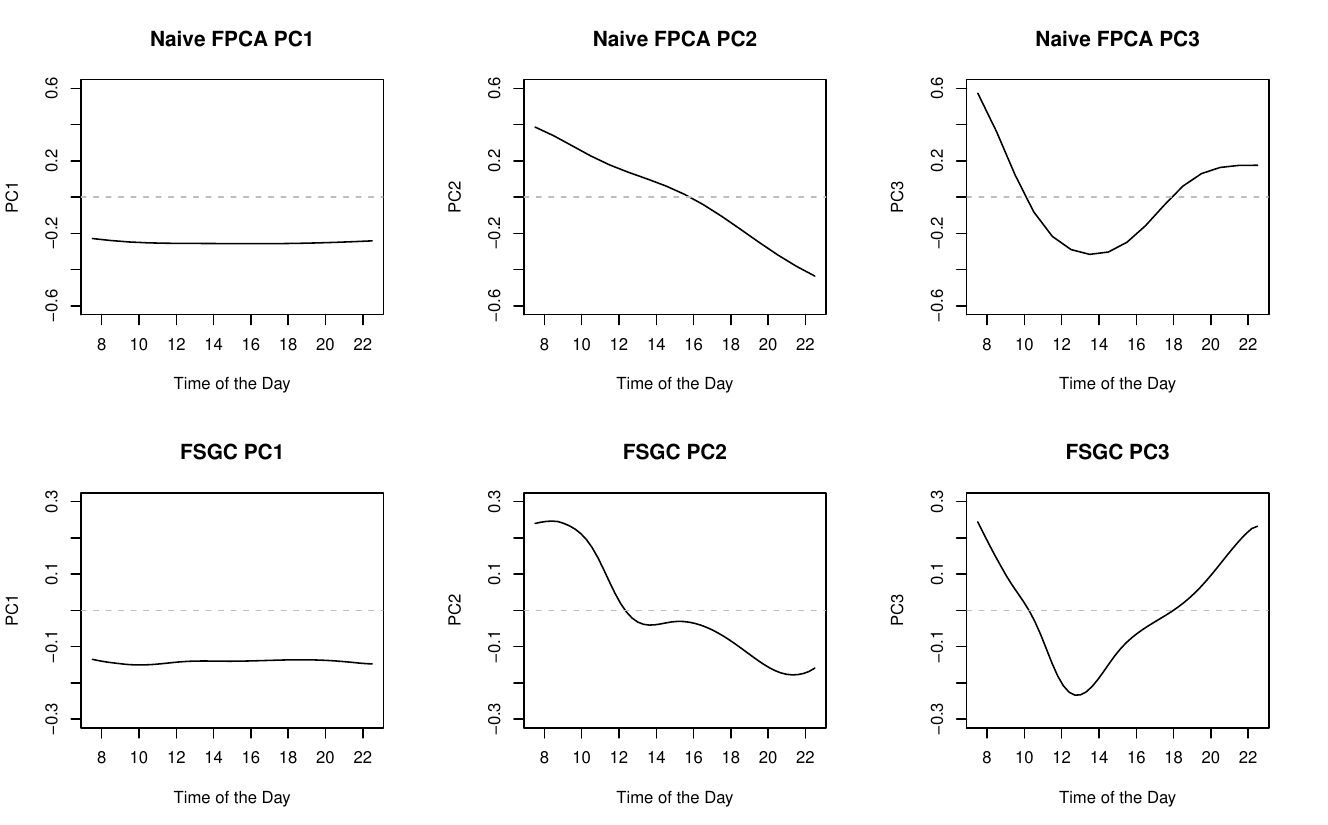}
\end{tabular}
\end{center}
\caption{Top panel shows estimated FSGC (left) and naive FPCA (right) correlation matrices. The middle and bottom panels show the first three naive FPCA and FSGC fPCs, respectively.}
\label{fig:fig3temp}
\end{figure}
Several pairs of time-combinations have a small number of observations, so we apply methods from Section 2.2 for sparse functional data. We use $K=5$ cubic B-spline basis functions with knots equally spaced between 7:30 AM and 10:30 PM. Our analysis proceeds as follows. First, we obtain and interpret estimated latent covariance and principal components. Second, we calculate predictions of latent representations of the observed data and use them to obtain latent principal component scores. Finally, we aggregate subject-specific scores into subject-specific mean and standard deviations of scores and associate those summaries with affective disorder diagnosis. 
The top panel of Figure \ref{fig:fig3temp} displays the estimated FSGC and naive FPCA estimated covariances. 

Note that the naive FPCA captures covariance between observed data and is shown here for illustrative purposes. Visually compared to naive FPCA, FSGC covariance demonstrates larger and more complex within-day temporal dependence and morning to early noon FSGC correlations (8 AM-2 PM) are higher. 
The bottom panel of Figure \ref{fig:fig3temp} shows the first three FSGC fPCs that capture approximately $66\%, 8\%,$ and $4\%$ of total variability, respectively. The middle panel of Figure \ref{fig:fig3temp} shows the first three naive fPCs that capture approximately $56\%, 10\%,$ and $3\%$ of total variability, respectively. Note that both approaches provide visually similar estimated fPCs. This is expected as the number of ordinal categories is relatively large, and we have a very large sample size. FSGC fPCs can be interpreted as follows. Interpreting scientifically, \textcolor{black}{fPC1 captures a temporally global daily mean of happiness/unhappiness; fPC2 estimates a contrast between the morning part of the day (7:30 AM-1 PM) and the rest of the day (1 PM-11 PM); fPC3 captures the degree of evening return to the morning happiness/unhappiness level. Hence, subjects with lower fPC1 scores (in absolute value) will be happier than subjects with lower scores; subjects with higher (positive) fPC2 scores will be less happy in the morning compared to the afternoon-late evening. Finally, subjects with higher (positive) fPC3 scores will have a more pronounced ``return" to the morning levels. It is interesting to note that fPC3 from naive FPCA has an evening return level significantly lower than the morning level. In contrast, fPC3 from FSGC has very similar morning and evening levels.} 

\begin{figure}[h]
     \centering
    \subfloat[Observed (ordinal)\label{fig:obs-dow}]{%
         \includegraphics[height=0.25\textwidth,width=0.48\textwidth]{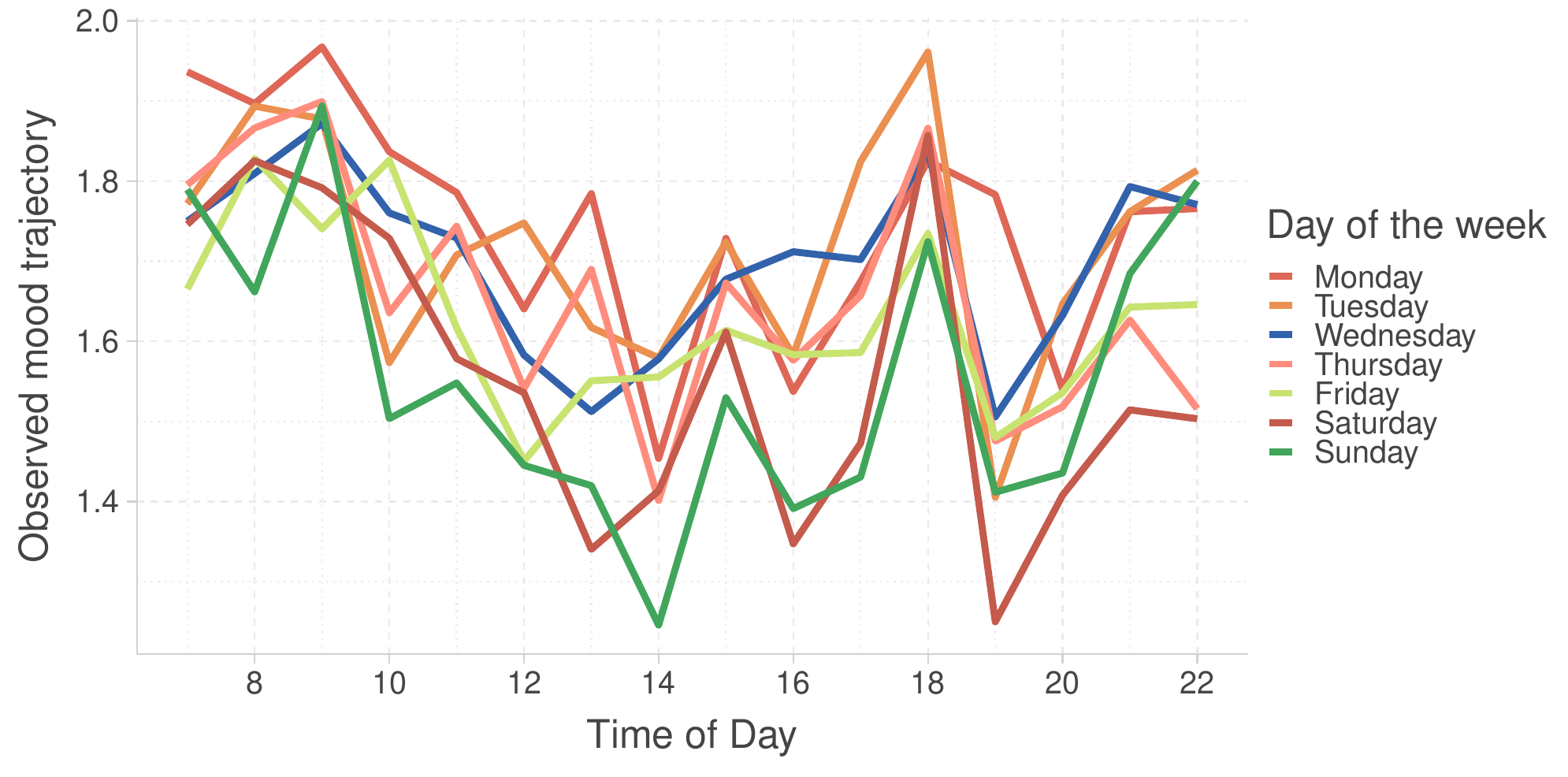}
         }
     \hfill
     \subfloat[Latent (continuous)\label{fig:lat-dow}]{%
         \includegraphics[height=0.25\textwidth,width=0.48\textwidth]{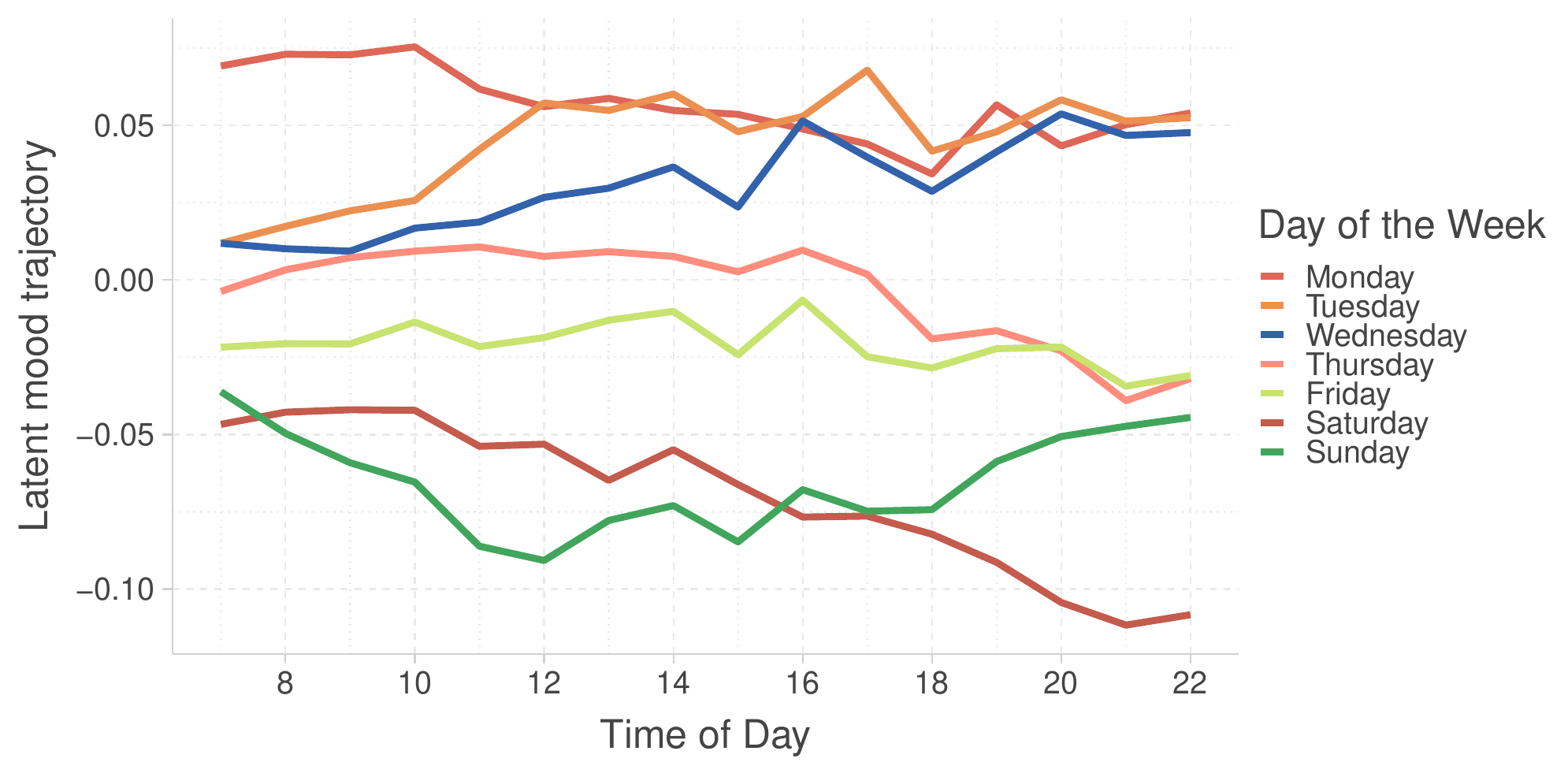}
         }
         \hfill
          \subfloat[Observed (ordinal)\label{fig:obs-diag}]{%
         \includegraphics[height=0.25\textwidth,width=0.45\textwidth]{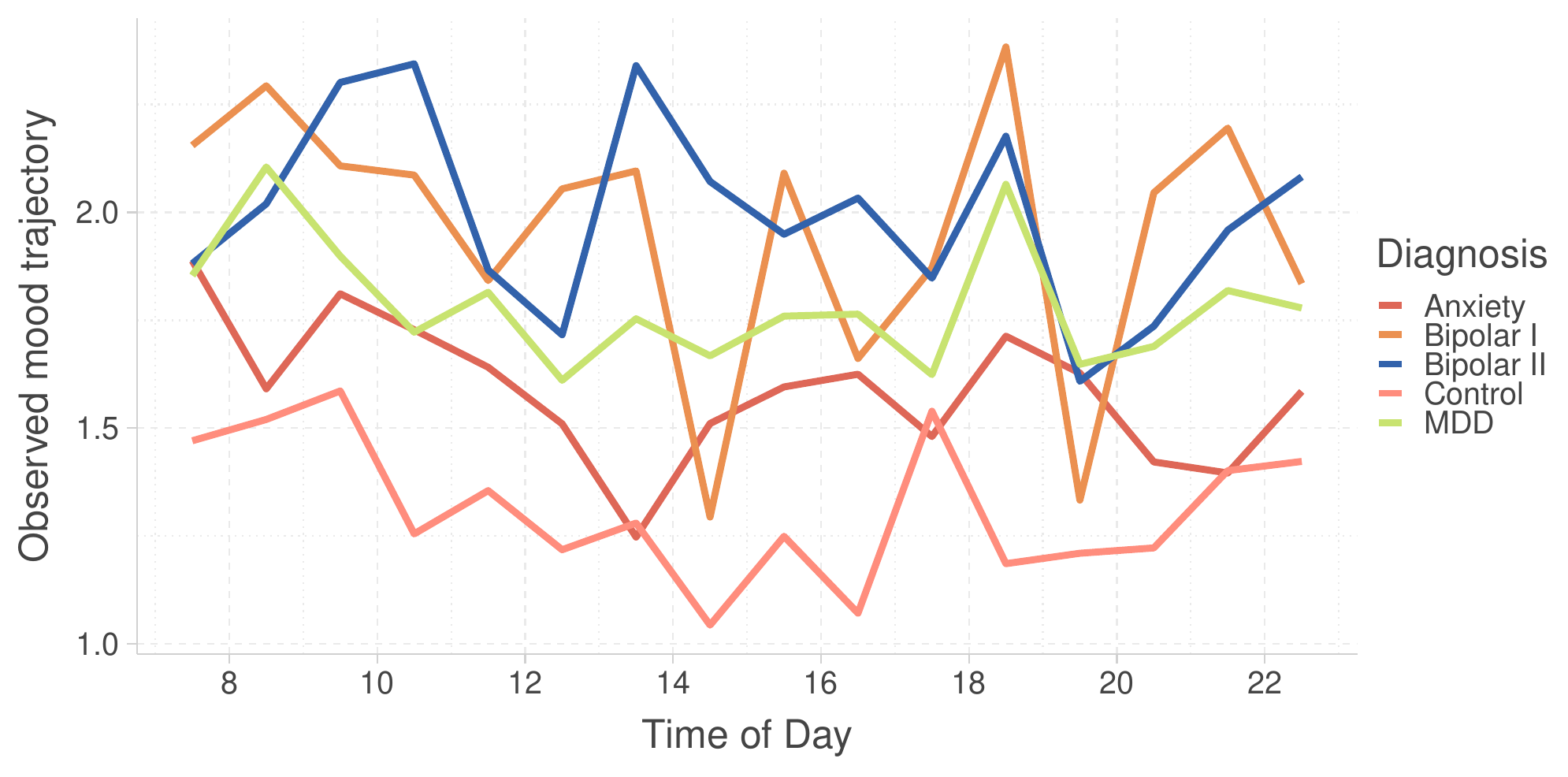}
         }
     \hfill
      \subfloat[Latent (continuous)\label{fig:lat-diag}]{%
         \includegraphics[height=0.25\textwidth,width=0.45\textwidth]{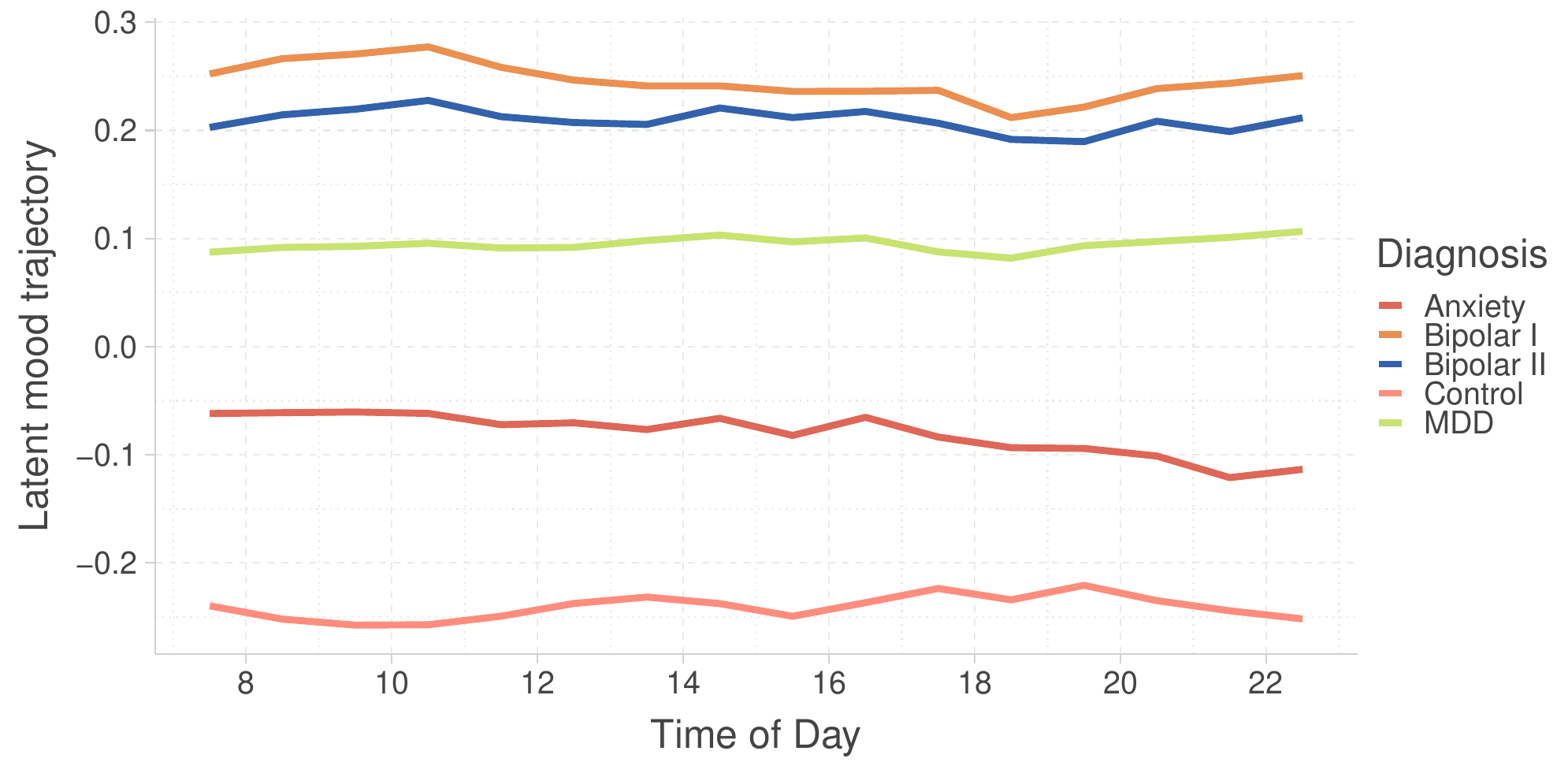}
         }
    \caption{Top panel shows within-day temporal patterns of mood averaged within each day of the week for observed ratings (left) and their latent representations (right). Bottom panel shows within-day temporal patterns of mood averaged within each mood disorder subtypes for observed ratings (left) and their latent representations (right). }
    \label{fig:mood-avg-comb}
\end{figure}

Another interesting application of our approach is estimating latent representations of reported mood and understanding time-of-day and day-of-week differences in their temporal patterns. We compare day-of-week average latent temporal patterns in the top panel of Figure \ref{fig:mood-avg-comb}. Our goal is to learn the effect of social schedules on emotional states. The latent trajectory reveals a clear clustering in the night-time levels of happiness: lowest in the early week (Monday to Wednesday), increases in the midweek (Thursday to Friday), and reaches highest in the weekend (Saturday to Sunday). This separation in the diurnal pattern of emotional states between days of the week is not visually apparent in the observed trajectories. 

One of the primary goal of this analysis is to explore the associations between within-day temporal patterns of mood disorder subtypes. Latent trajectories averaged within each diagnosis (Figure \ref{fig:mood-avg-comb} bottom panel), reveal an ordering of unhappiness between the diseases: Bipolar I $>$ Bipolar II $>$ MDD $>$ Anxiety $>$ Control. Within-day temporal trajectories are flatter than observed ones, but there is a much clearer separation between diagnosis groups. We next calculate the latent subject-day mood scores and aggregate them using average and standard deviation for each subject across days. 

 This gives us three mean and three SD latent FPC scores for each subject. Supplementary Figure S25 top and bottom panel displays the distribution of the latent mean FPC scores and SD FPC scores, respectively, for five mood-disorder groups. It appears that bipolar I, bipolar II, and MDD groups have lower mean fPC-1 scores compared to the controls. Anxiety and MDD groups seem to have higher SD fPC-1 scores than controls. Similarly, Bipolar II is observed to have a higher SD fPC-2 score compared to controls. Supplementary Figure S26 displays the correlation between mean and SD fPC scores. SD fPC scores are moderately positively correlated. We fit a multinomial logistic regression model for mood disorders subtypes with controls being the reference and mean and SD fPC scores as predictors while adjusting for age and sex (with male being the reference). Since SD fPC scores are correlated, we include them one at a time along with the mean FPC scores in competing models and choose the one with the lowest AIC. Table \ref{tab:my-table2} displays the estimated regression coefficients along their associated p-values (Wald test) for the chosen multinomial logistic regression model of affective disorders on the mean of fPC1, fPC2, fPC3, and SD of fPC2 scores. 
We use $\alpha=0.05$ for the statistical significance of all the tests.

A higher mean fPC1 score is found to be associated with a lower odds of bipolar I, bipolar II disorders, and MDD relative to controls. This clearly demonstrates that these diagnosis groups report different levels of daily mood. A higher SD fPC2 score, capturing the variability of the fPC2 score across days, is found to be associated with higher odds of bipolar II and MDD disorders relative to controls. 


\begin{table}[ht]
\hspace{- 50 mm}
\small
\caption{Estimated regression coefficients from multinomial logistic regression of mood disorders (Type) on the mean of the first three FSGC fPCs and the SD of FSGC fPC-2 scores, age and sex. (p-values are reported in the parenthesis)}
\label{tab:my-table2}
\resizebox{\textwidth}{!}{\begin{tabular}{|c|c|c|c|c|c|c|c|}
\hline
Type & Intercept     & Age           & sex (Female)  & mean FPC-1           & mean FPC-2           & mean FPC-3           & SD FPC-2  \\ \hline
Anxiety            & 0.01 (0.982)  & -0.024 (0.001) & 0.84 (0.003)  & -0.38 (0.094) & -2.82 (0.187) & -5.11 (0.126) & 1.27 (0.562) \\ \hline
bipolar I          & -1.36 (0.033) & -0.0002 (0.986)  & 0.88 (0.011)  & -1.41 ($<0.001$) & -2.94 (0.246)  & -1.84 (0.646) & -0.24 (0.931)\\ \hline
bipolar II         & -1.96 (0.003) & -0.014 (0.120) & 0.53 (0.123)  & -1.26 ($<0.001$) & 0.29 (0.906) & -5.06 (0.197) & 7.74 (0.004) \\ \hline
MDD                & -1.36 (0.005) & 0.002 (0.708)  & 1.10 ($<0.001$)  & -0.94 ($<0.001$) & -0.18 (0.923) & -3.03 (0.310) & 4.84 (0.016) \\ \hline
\end{tabular}}
\end{table}
The proposed FSGC method provides novel insights into the within-day temporal patterns of mood and their association with different mood disorder subtypes.

\section{Discussion}
In this article, we have developed a unified functional principal component analysis for continuous non-Gaussian, truncated, ordinal, and binary functional data using Semiparametric Gaussian Copula. The method is rank-based that results in invariance to monotone transformation of scale and robustness. Numerical simulations have illustrated the satisfactory performance of the proposed method for binary, ordinal, truncated, and continuous functional data under dense and sparse sampling designs compared to existing covariance estimation methods. In the application to the mHealth NIMH study, FSGC identified principal within-day temporal patterns of mood  associated with mood disorder subtypes.

The proposed covariance estimation is built on a tensor product spline representation of the covariance of the latent process. The choice of the basis functions (B-splines) controls the smoothness of the estimated covariance via a truncated basis approach. The choice of the number of basis functions is subjective and depends on the design and the number of time points for each subject. We recommend using a moderate number of basis functions based on the number of unique time points. A data-driven choice for this could be developed using cross-validated curve predictions, as illustrated in Section 2.3. Alternatively, a roughness penalty can be incorporated in the objective function \citep{Ramsay05functionaldata,xiao2016fast}. This would result in a penalized nonlinear least square problem. We leave this approach and choice of smoothing penalty as a direction to be explored more deeply in the future.
\textcolor{black}{In this article, we have followed a point-wise estimation strategy for cutoff  $\bDelta(t)$ and transformation $f_t$ functions (for continuous and truncated data). Since no explicit smoothness is enforced, this formulation allows for very flexible models, including rough, non-continuous observed processes $X(t)$. If desired, smoothness can be enforced in $\bDelta(t)$ and $f_t(\cdot)$ by, first, estimating this point-wise and, then, using splines or kernel-based smoothers.}

Multiple research directions remain to be explored. We have focused on modeling univariate functional data of the same type in this article. Alternatively, modeling multivariate mixed-type functional data would be an important next step that would model the joint latent dependence between different mixed types \citep{dey2022semiparametric} scales such as physical activity (continuous) and pain (truncated) or pain (truncated) and mood (ordinal). This step is crucial since building separate marginal models and ignoring the cross-correlation between different variables leads to information loss that might result in biased predictions \citep{li2014hierarchical}. Following \cite{di2009multilevel}, multilevel extension of FSGC can be developed to consider multi-level designs. Additionally, FSGC is primarily developed for dense and moderately sparse designs. The approach can be extended for highly sparse and irregular functional data using kernel-weighted objective functions that borrow information from neighboring observations. Extending the proposed method to such general class models would allow more diverse applications and remain areas for future research based on this current work.

\section{Software}

A \verb|R| package implementation of the FSGC estimation has been made available as \verb|fpca.sgc.lat| function as part of the \verb|SGCTools| package (\url{https://github.com/Ddey07/SGCTools}). We also present a reproducible \verb|RMarkdown| demonstrating the use of the package in \verb|Github|. 






\section*{Supplementary Material}
Supplementary Tables S1-S2 and Supplementary Figures S1-S27 referenced in this article are available online as Supplementary Material.


\bibliographystyle{jasa3} 
\bibliography{lit.bib}



\newpage 

\begin{center}
{\large\bf Supplementary Materials}
\end{center}

\beginsupplement

\section{Appendix A : Analytic forms of bridging functions}\label{appn:bridge}

We denote $\rho_{jj'} = C(t_j, t_{j'})$ and the suffices of $F$ - \{cc, tt, oo, bb\} denote specific cases of continuous, truncated, ordinal and binary variables respectively. 

\begin{align}
     F_{\rm cc}(\rho_{jj'}) & = \dfrac{2}{\pi} \sin^{-1}(\rho_{jj'}) \nonumber\\
F_{\rm bb}(\rho_{jj'}; \Delta(t_j), \Delta(t_j'))&  = 2\left\{\Phi_2( \Delta(t_j),  \Delta(t_{j'}); \rho_{jj'})-\Phi( \Delta(t_j))\Phi( \Delta(t_{j'}))\right\} \nonumber\\
F_{\rm tt}(\rho_{jj'}; \Delta(t_j), \Delta(t_{j'}))  & = -2 \Phi_4 (-\Delta(t_j), -\Delta(t_{j'}), 0,0; S_{4a}(\rho_{jj'})) + 2 \Phi_4 (-\Delta(t_j), -\Delta(t_{j'}), 0,0; S_{4b}(\rho_{jj'})) \nonumber\\
F_{\rm oo}(\rho_{jj'}; \Delta(t_j), \Delta(t_{j'})) &= 2 \sum_{r=1}^{l_j-1}\sum_{s=1}^{l_{j'}-1}[\Phi_2(\Delta_{r}(t_j),\Delta_{s}(t_{j'}); \rho_{jj'})\{ \Phi_2(\Delta_{(r+1)}(t_j),\Delta_{(s+1)}(t_{j'});\rho_{jj'}) - \nonumber \\
& \Phi_2(\Delta_{(r+1)}(t_j),\Delta_{(s-1)}(t_{j'}; \rho_{jj'})\}] - 2 \sum_{r=1}^{l_j-1}\Phi({\Delta_{r}(t_j)})\Phi_2(\Delta_{(r+1)}(t_j),\Delta_{(l_{j'}-1)(t_{j'})}; \rho_{jj'}) 
\end{align}
with 
\begin{equation}
\begin{split}\nonumber
S_{4b}(\rho_{jj'}) & = \bpm
    1 & \rho_{jj'} & 1/\sqrt{2} & \rho_{jj'}/\sqrt{2}\\
    \rho_{jj'} & 1 & \rho_{jj'}/\sqrt{2}& 1/\sqrt{2}\\
    1/\sqrt{2}& \rho_{jj'}/\sqrt{2} & 1& \rho_{jj'} \\ \rho_{jj'}/\sqrt{2}& 1/\sqrt{2}& \rho_{jj'}  & 1
\epm
\end{split}
\end{equation}

\section{Appendix B: Additional Simulation Scenario Results}
The results from the additional simulation scenario (Scenario A, sparse design) are reported in Web Figure 5-6. In this sparse case the bfpca method is no longer applicable. We observe that the proposed FSGC method outperforms the FSGC latent approach in this case, illustrating the applicability of the proposed approach in sparse designs. For the sparse scenario, we also apply the proposed curve prediction method in Section 2.3 of the paper and obtain $\hat{X_i}(t)$ at all time points over $\mathcal{S}$ (a dense grid on $\mathcal{T}$). We calculate the accuracy of these binary predictions by comparing them with the true curves $X_i(t)$ across all time points and all subjects. While calculating accuracy, we ignore the points $S_i$, where the curve $X_i(\cdot)$ was already observed. For a particular M.C replication this is calculated as $ACC=\frac{1}{n}\sum_{i=1}^n\frac{1}{m-n_i}\sum_{t_j\in \mathcal{S}-S_i}I(X_i(t_j)=\hat{X}_i(t_j))$. The average accuracy across all M.C replications
is found to be 0.98 ($sd=0.002$) and 0.95 ($sd=0.002$) for the stationary and non-stationary covariance scenario, respectively, illustrating the satisfactory performance of the proposed method in predicting the binary curves at new-time points.

\section{Supplementary Figures}

\subsection*{Binary Functional Data}
\begin{figure}[H]
\centering
\includegraphics[width=.9\linewidth , height=1\linewidth]{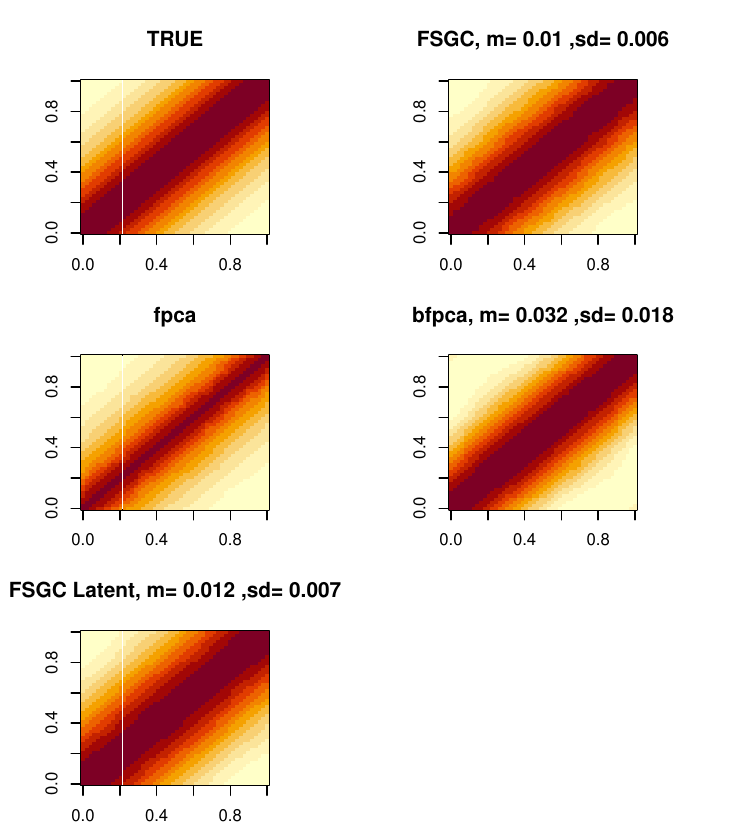}
\caption{True and Estimated average covariance surface for stationary covariance kernel, scenario A, n=100. The average ISE (and sd) of the estimates are reported on the top of the respective images. FSGC denotes the proposed estimation method, fpca denotes FPCA on the observed curve, bfpca denotes estimate from binary FPCA and FSGC Latent is FPCA on latent predictions from SGCRM. }
\label{fig:fig1-s}
\end{figure}

\begin{figure}[H]
\centering
\includegraphics[width=.9\linewidth , height=1\linewidth]{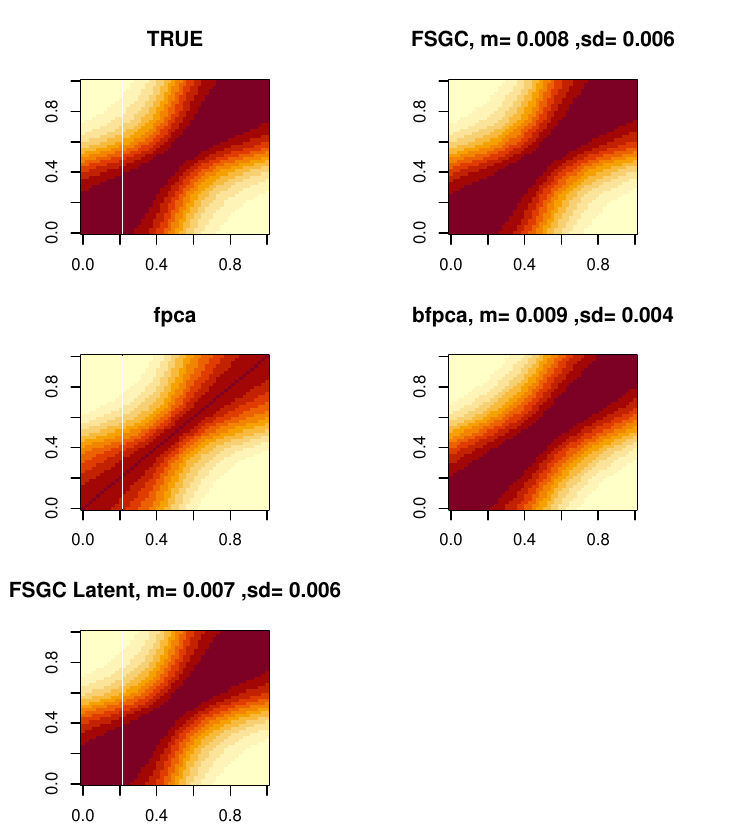}
\caption{True and Estimated average covariance surface for non-stationary covariance kernel, scenario A, n=100. Average ISE (and sd) of the estimates are reported on the top of the respective images. FSGC denotes the proposed estimation method, fpca denotes FPCA on the observed curve, bfpca denotes estimate from binary FPCA and FSGC Latent is FPCA on latent predictions from SGCRM. }
\label{fig:fig2-s}
\end{figure}

\begin{figure}[H]
\centering
\includegraphics[width=.9\linewidth , height=0.93\linewidth]{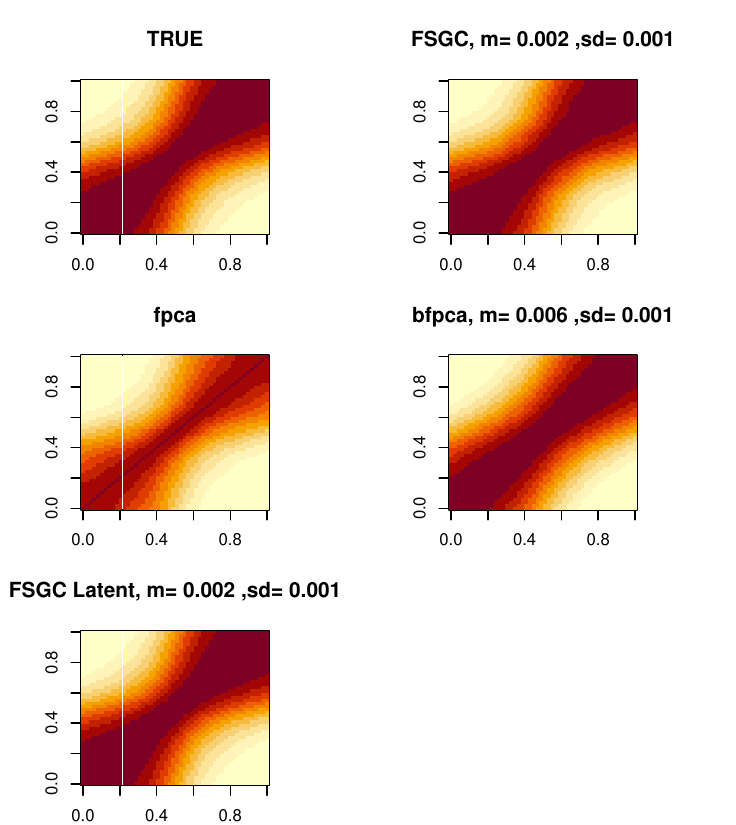}
\caption{True and estimated average covariance surface for non-stationary covariance kernel, scenario A, n=500. The average ISE (and sd) of the estimates are reported on the top of each image. FSGC denotes the proposed FSGC method, fpca denotes naive FPCA of the observed data, bfpca denotes binary FPCA, and FSGC Latent denotes FPCA on latent predictions from SGCRM.}
\label{fig:fig2}
\end{figure}

\begin{figure}[H]
\centering
\includegraphics[width=.9\linewidth , height=1\linewidth]{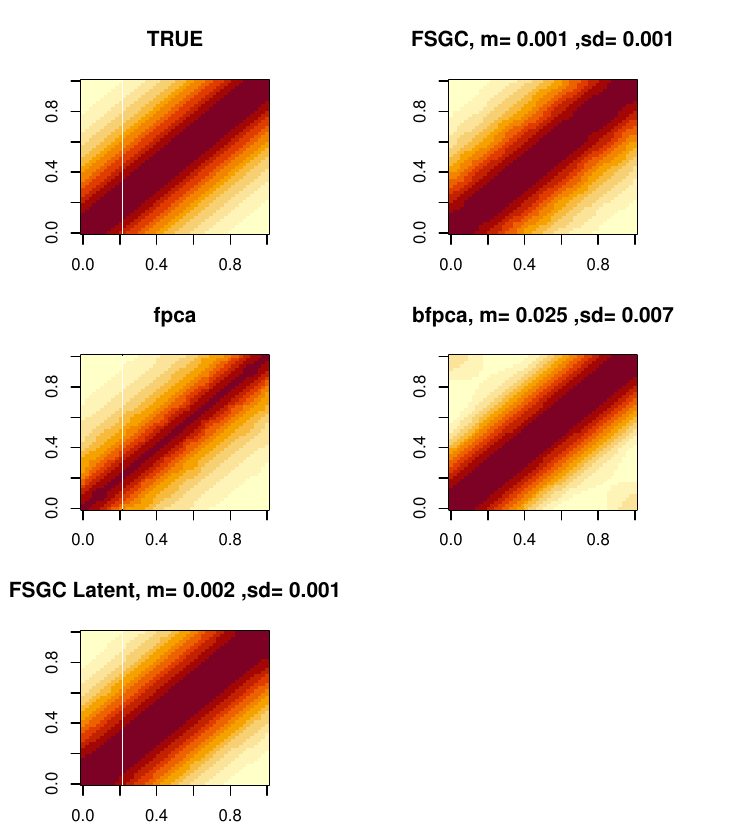}
\caption{True and Estimated average covariance surface for stationary covariance kernel, scenario A, n=1000. Average ISE (and sd) of the estimates are reported on the top of the respective images. FSGC denotes the proposed estimation method, fpca denotes FPCA on the observed curve, bfpca denotes estimate from binary FPCA and FSGC Latent is FPCA on latent predictions from SGCRM. }
\label{fig:fig3}
\end{figure}

\begin{figure}[H]
\centering
\includegraphics[width=.9\linewidth , height=1\linewidth]{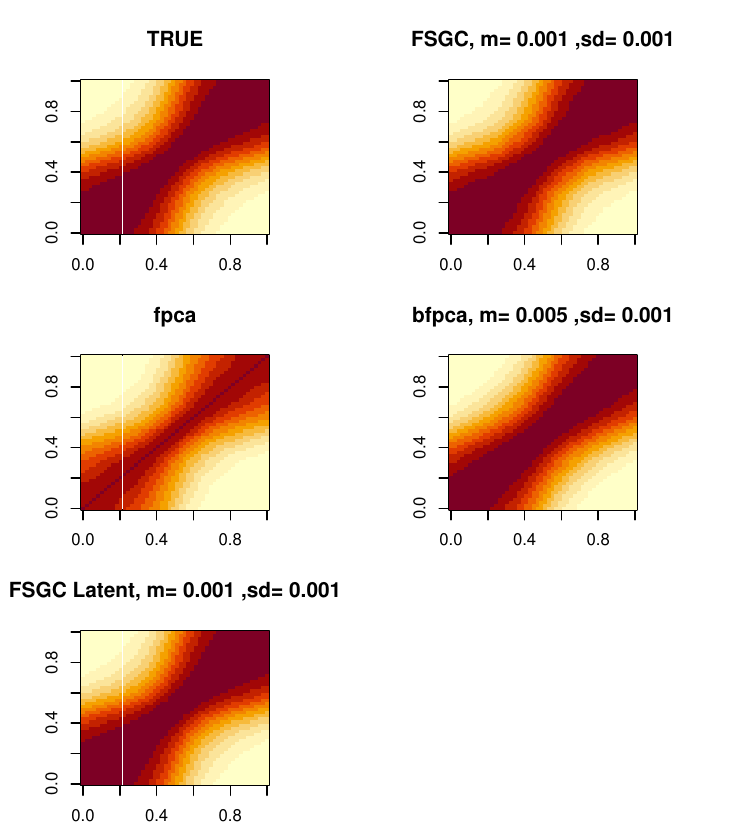}
\caption{True and Estimated average covariance surface for non-stationary covariance kernel, scenario A, n=1000. Average ISE (and sd) of the estimates are reported on the top of the respective images. FSGC denotes the proposed estimation method, fpca denotes FPCA on the observed curve, bfpca denotes estimate from binary FPCA and FSGC Latent is FPCA on latent predictions from SGCRM. }
\label{fig:fig4}
\end{figure}

\begin{figure}[H]
\centering
\includegraphics[width=1\linewidth , height=.8\linewidth]{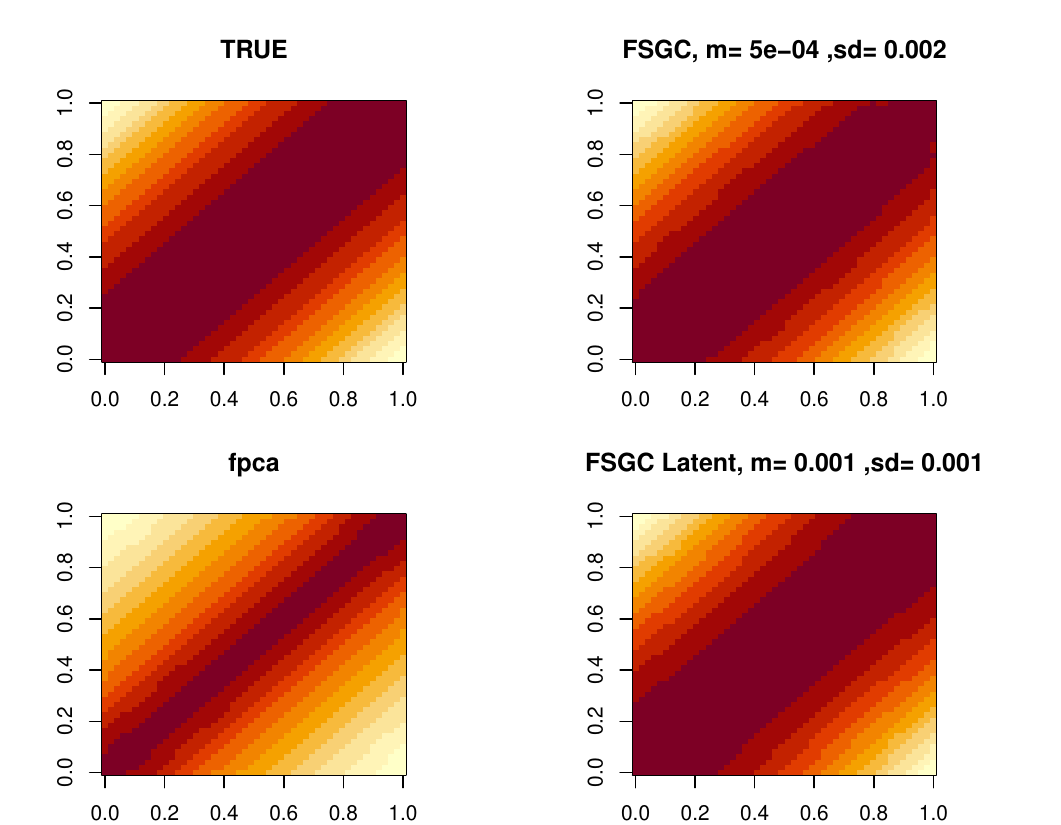}
\caption{True and Estimated average covariance surface for stationary covariance kernel, sparse design, scenario A, n=1000. Average ISE (and sd) of the estimates are reported on the top of the respective images. FSGC denotes the proposed estimation method, fpca denotes FPCA on the observed curve and FSGC Latent is FPCA on latent predictions from SGCRM. }
\label{fig:fig5}
\end{figure}

\begin{figure}[H]
\centering
\includegraphics[width=1\linewidth , height=0.8\linewidth]{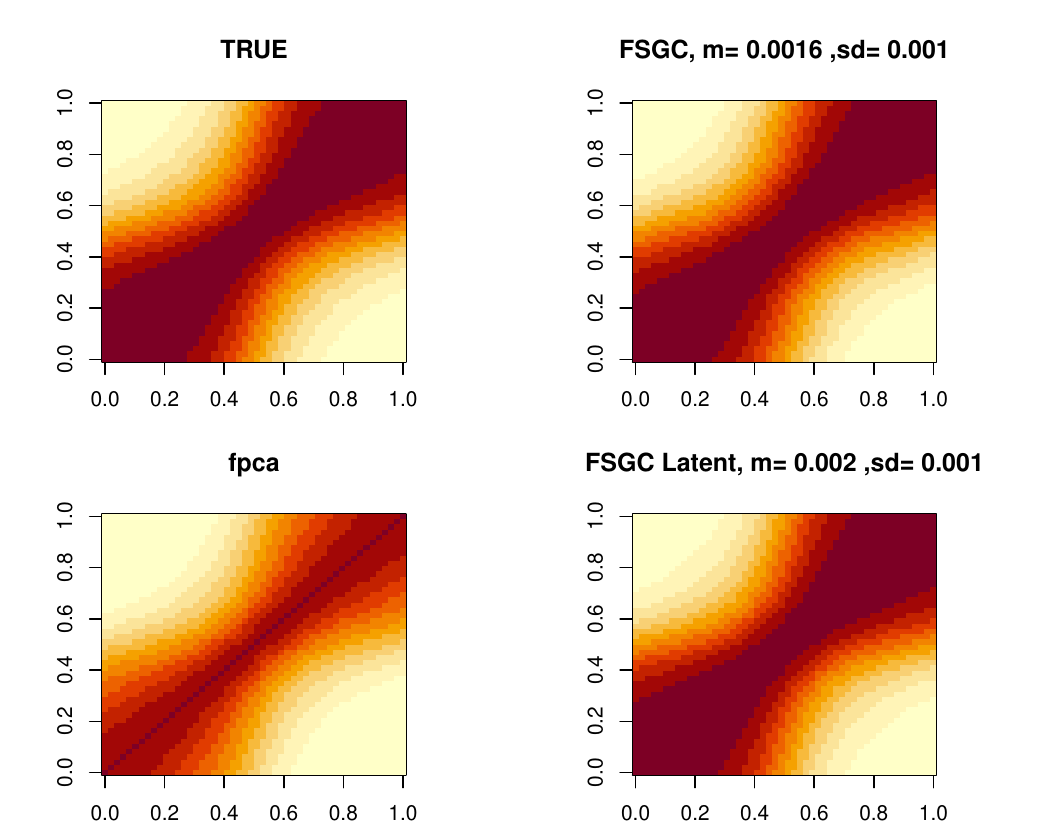}
\caption{True and Estimated average covariance surface for non-stationary covariance kernel, sparse design, scenario A, n=1000. Average ISE (and sd) of the estimates are reported on the top of the respective images. FSGC denotes the proposed estimation method, fpca denotes FPCA on the observed curve and FSGC Latent is FPCA on latent predictions from SGCRM. }
\label{fig:fig6}
\end{figure}

\subsection*{Ordinal Functional Data}

\begin{figure}[H]
\centering
\includegraphics[width=1\linewidth , height=0.9\linewidth]{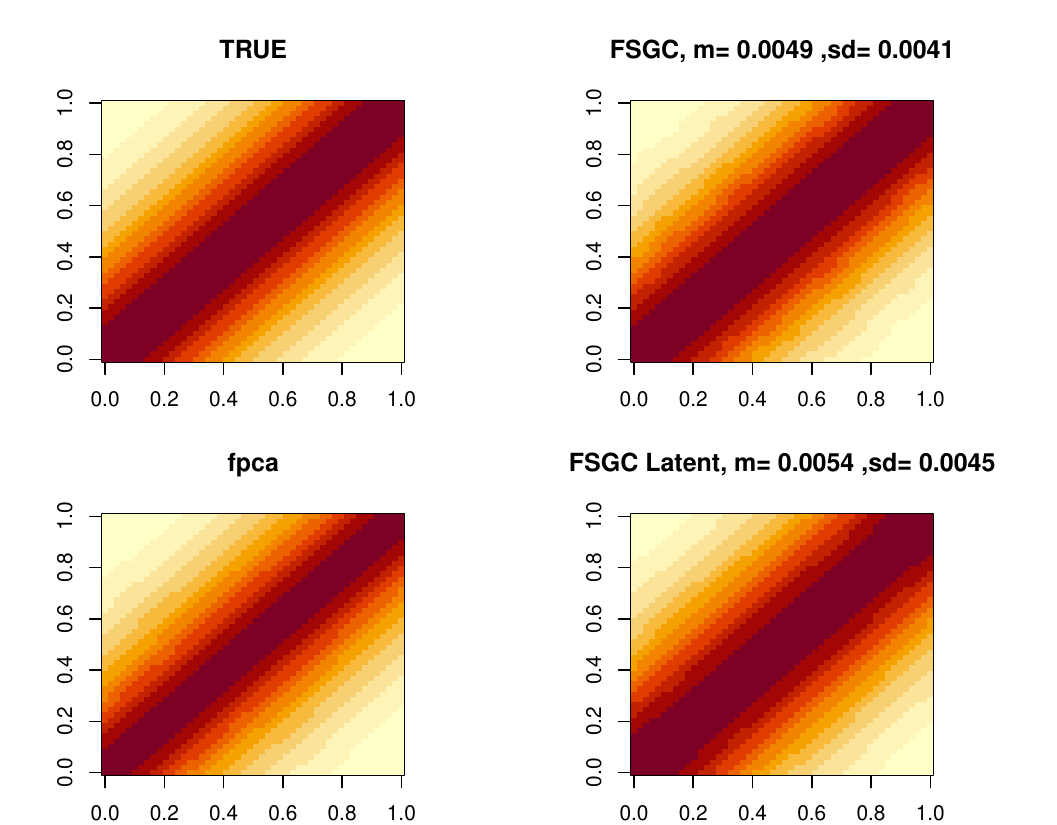}
\caption{True and Estimated average covariance surface for stationary covariance kernel, scenario B, n=100. Average ISE (and sd) of the estimates are reported on the top of the respective images. FSGC denotes the proposed estimation method, fpca denotes FPCA on the observed curve and FSGC Latent is FPCA on latent predictions from SGCRM. }
\label{fig:fig7}
\end{figure}
\begin{figure}[H]
\centering
\includegraphics[width=1\linewidth , height=0.9\linewidth]{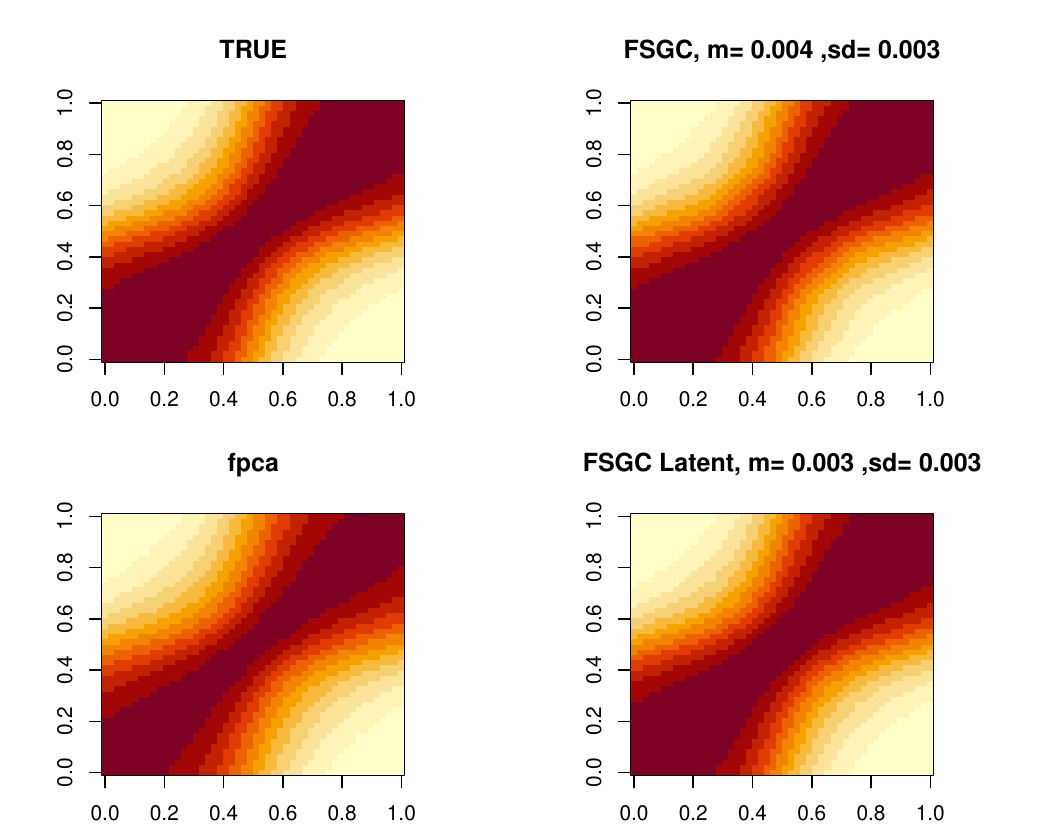}
\caption{True and Estimated average covariance surface for non-stationary covariance kernel, scenario B, n=100. Average ISE (and sd) of the estimates are reported on the top of the respective images. FSGC denotes the proposed estimation method, fpca denotes FPCA on the observed curve and FSGC Latent is FPCA on latent predictions from SGCRM. }
\label{fig:fig8}
\end{figure}

\begin{figure}[H]
\centering
\includegraphics[width=1\linewidth , height=0.9\linewidth]{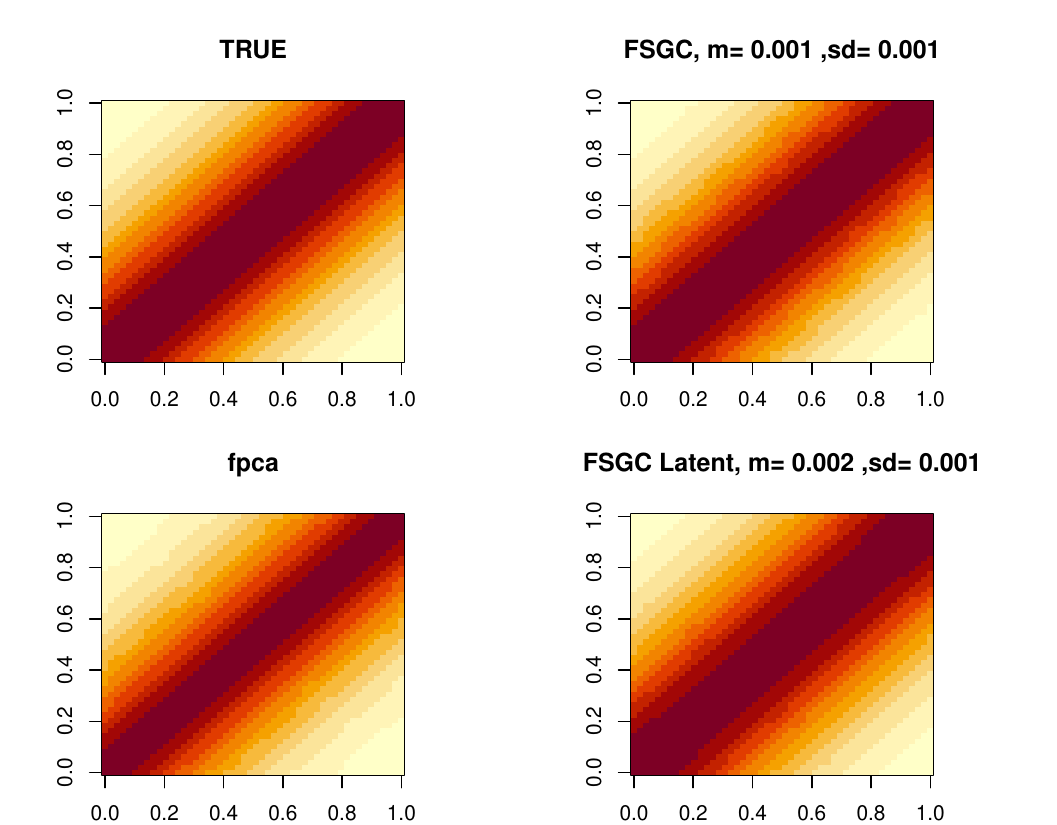}
\caption{True and Estimated average covariance surface for stationary covariance kernel, scenario B, n=500. Average ISE (and sd) of the estimates are reported on the top of the respective images. FSGC denotes the proposed estimation method, fpca denotes FPCA on the observed curve and FSGC Latent is FPCA on latent predictions from SGCRM. }
\label{fig:fig9}
\end{figure}

\begin{figure}[H]
\centering
\includegraphics[width=1\linewidth , height=0.9\linewidth]{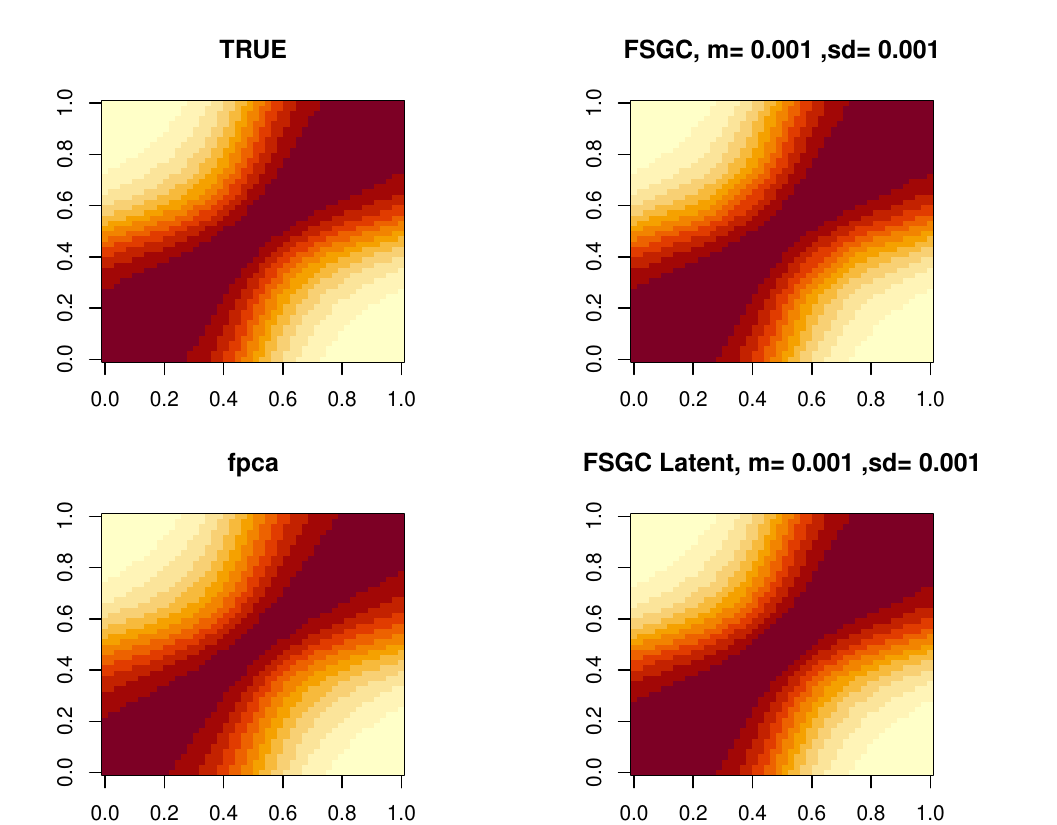}
\caption{True and Estimated average covariance surface for non-stationary covariance kernel, scenario B, n=500. Average ISE (and sd) of the estimates are reported on the top of the respective images. FSGC denotes the proposed estimation method, fpca denotes FPCA on the observed curve and FSGC Latent is FPCA on latent predictions from SGCRM. }
\label{fig:fig10}
\end{figure}

\begin{figure}[H]
\centering
\includegraphics[width=1\linewidth , height=0.9\linewidth]{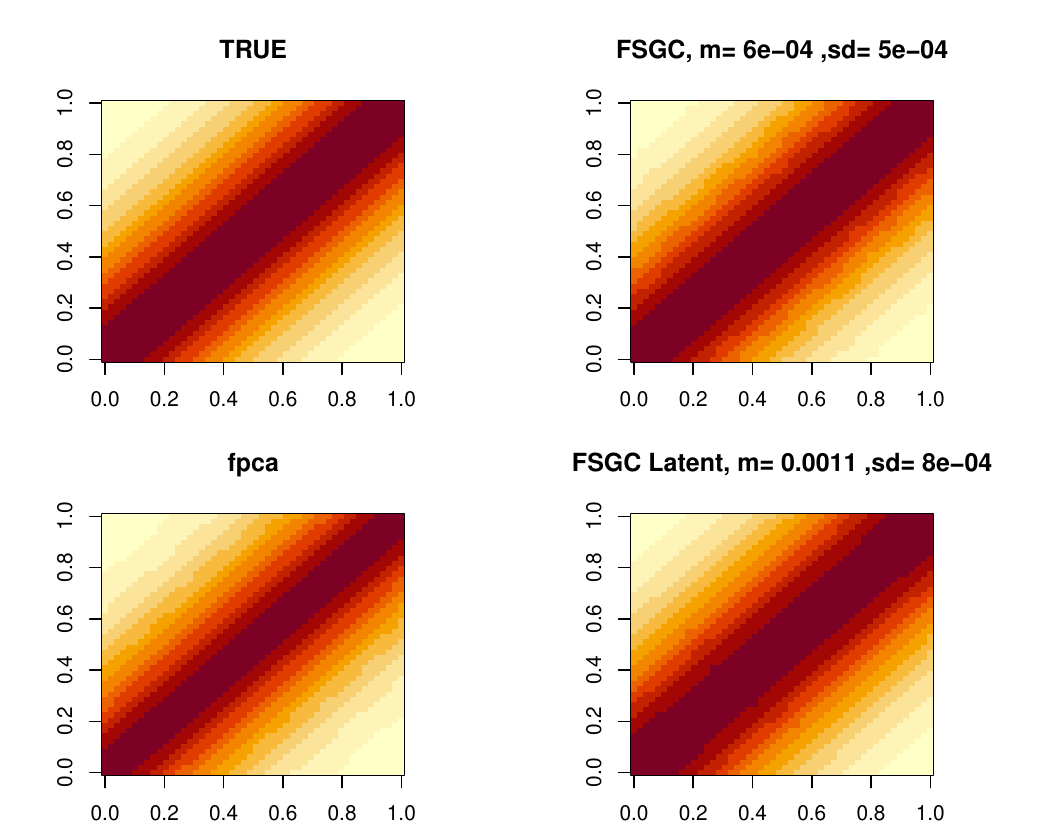}
\caption{True and Estimated average covariance surface for stationary covariance kernel, scenario B, n=1000. Average ISE (and sd) of the estimates are reported on the top of the respective images. FSGC denotes the proposed estimation method, fpca denotes FPCA on the observed curve and FSGC Latent is FPCA on latent predictions from SGCRM. }
\label{fig:fig11}
\end{figure}

\begin{figure}[H]
\centering
\includegraphics[width=1\linewidth , height=0.9\linewidth]{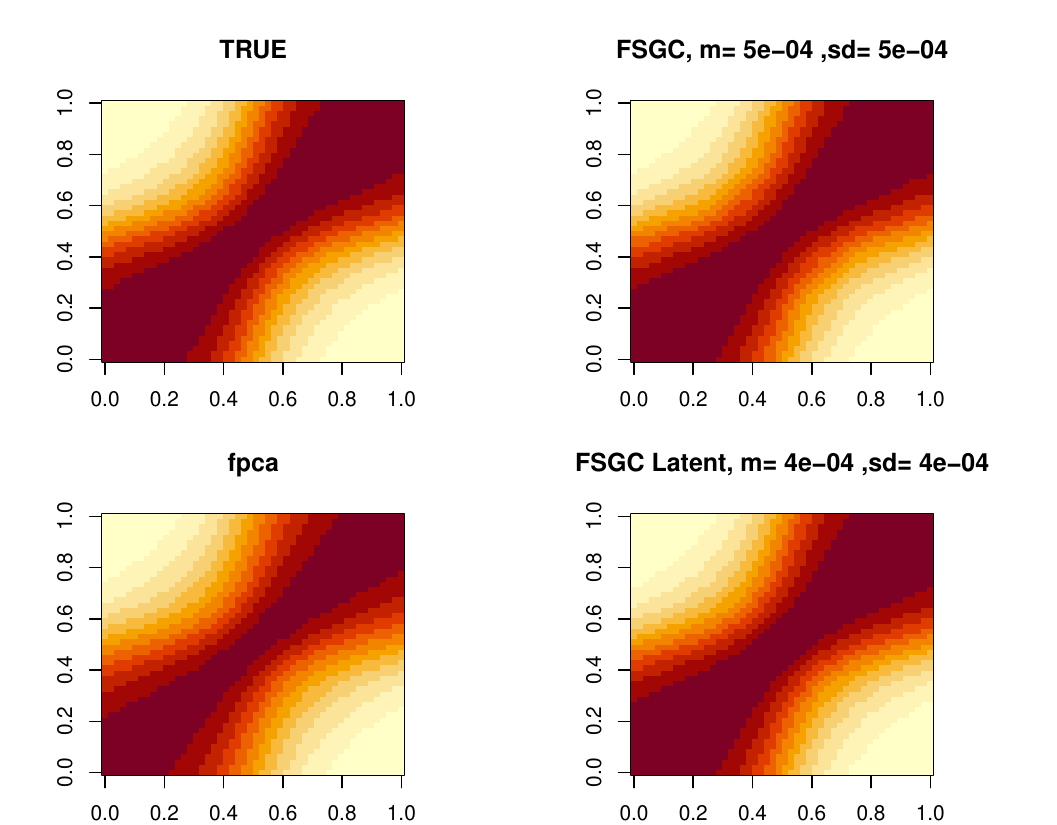}
\caption{True and Estimated average covariance surface for non-stationary covariance kernel, scenario B, n=1000. Average ISE (and sd) of the estimates are reported on the top of the respective images. FSGC denotes the proposed estimation method, fpca denotes FPCA on the observed curve and FSGC Latent is FPCA on latent predictions from SGCRM. }
\label{fig:fig12}
\end{figure}

\subsection*{Truncated Functional Data}
\begin{figure}[H]
\centering
\includegraphics[width=1\linewidth , height=0.9\linewidth]{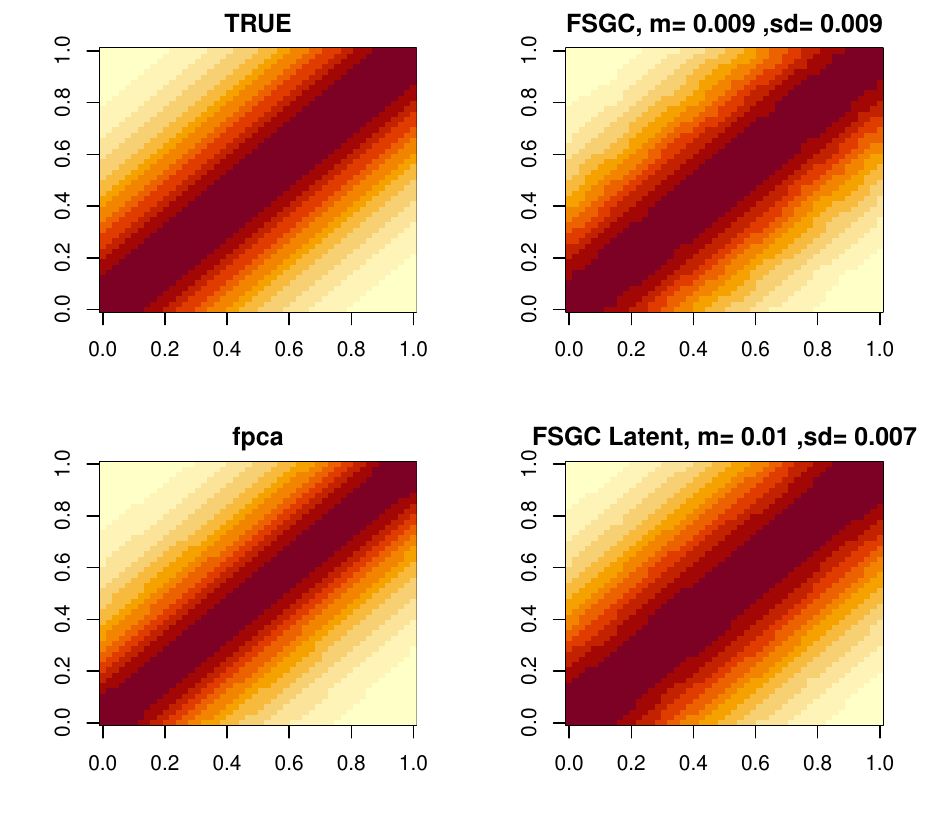}
\caption{True and Estimated average covariance surface for stationary covariance kernel, scenario C, n=100. Average ISE (and sd) of the estimates are reported on the top of the respective images. FSGC denotes the proposed estimation method, fpca denotes FPCA on the observed curve and FSGC Latent is FPCA on latent predictions from SGCRM. }
\label{fig:fig13}
\end{figure}
\begin{figure}[H]
\centering
\includegraphics[width=1\linewidth , height=0.9\linewidth]{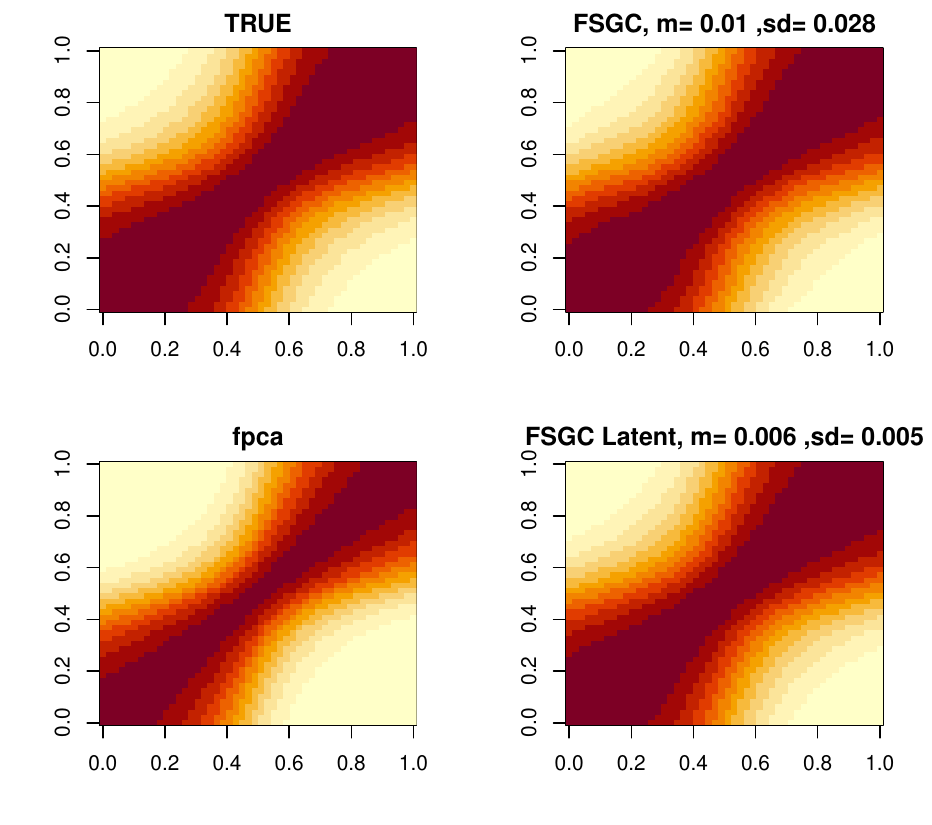}
\caption{True and Estimated average covariance surface for non-stationary covariance kernel, scenario C, n=100. Average ISE (and sd) of the estimates are reported on the top of the respective images. FSGC denotes the proposed estimation method, fpca denotes FPCA on the observed curve and FSGC Latent is FPCA on latent predictions from SGCRM. }
\label{fig:fig14-tr}
\end{figure}

\begin{figure}[H]
\centering
\includegraphics[width=1\linewidth , height=0.9\linewidth]{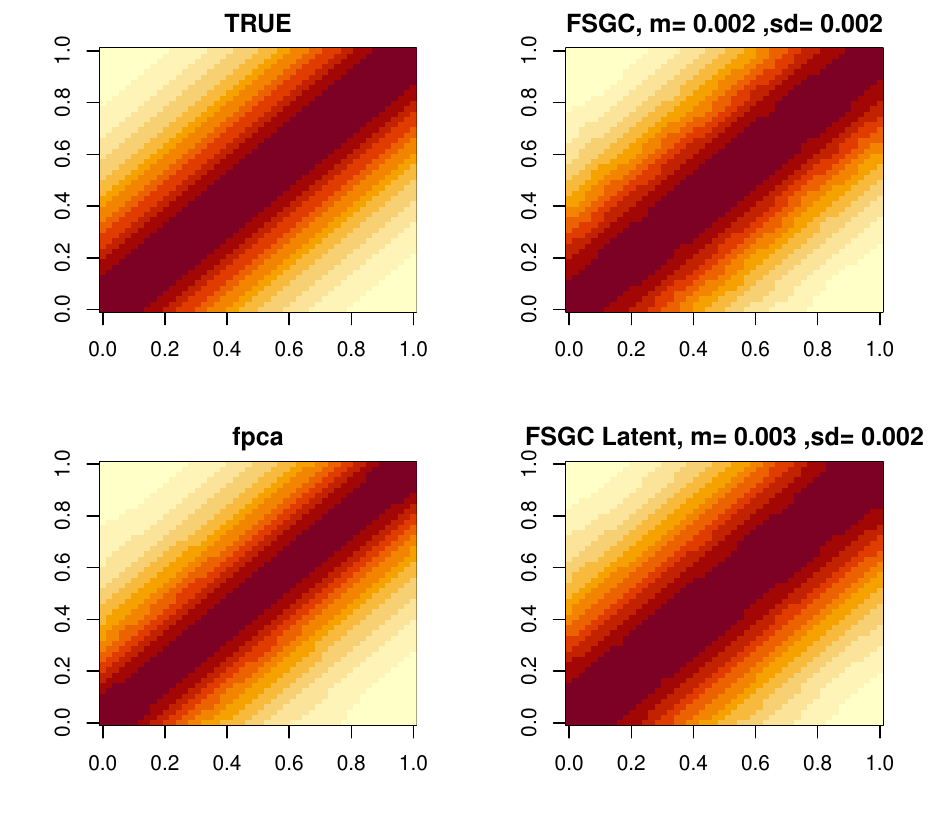}
\caption{True and Estimated average covariance surface for stationary covariance kernel, scenario C, n=500. Average ISE (and sd) of the estimates are reported on the top of the respective images. FSGC denotes the proposed estimation method, fpca denotes FPCA on the observed curve and FSGC Latent is FPCA on latent predictions from SGCRM. }
\label{fig:fig15}
\end{figure}

\begin{figure}[H]
\centering
\includegraphics[width=1\linewidth , height=0.9\linewidth]{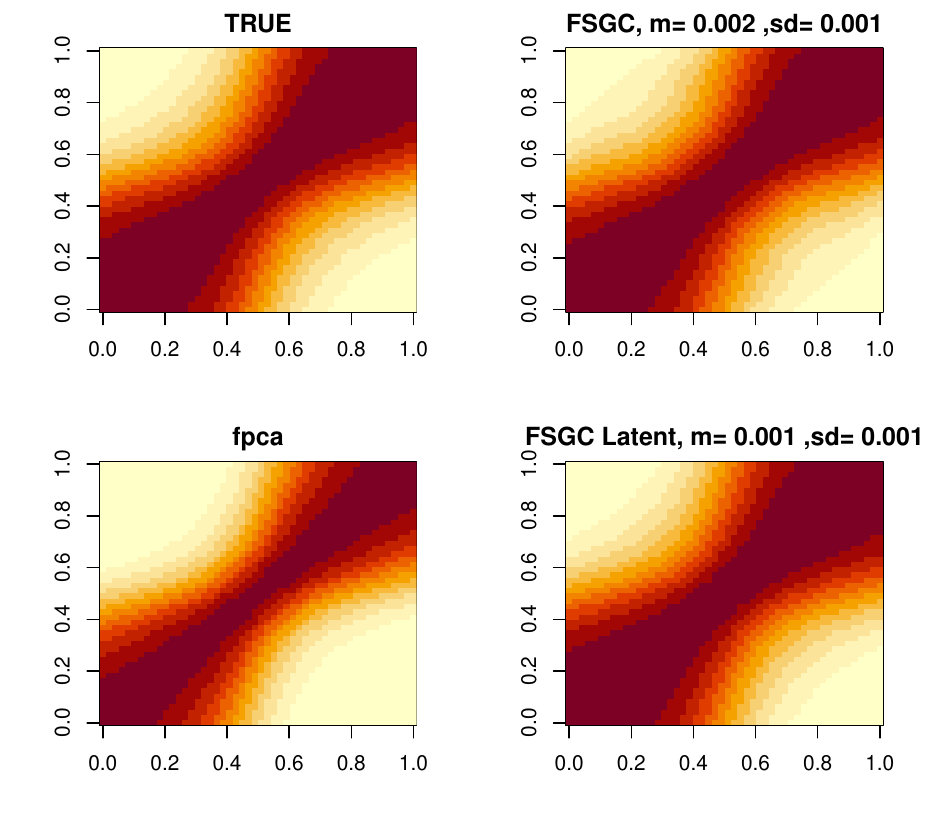}
\caption{True and Estimated average covariance surface for non-stationary covariance kernel, scenario C, n=500. Average ISE (and sd) of the estimates are reported on the top of the respective images. FSGC denotes the proposed estimation method, fpca denotes FPCA on the observed curve and FSGC Latent is FPCA on latent predictions from SGCRM. 
}
\label{fig:fig16-tr}
\end{figure}

\begin{figure}[H]
\centering
\includegraphics[width=1\linewidth , height=0.9\linewidth]{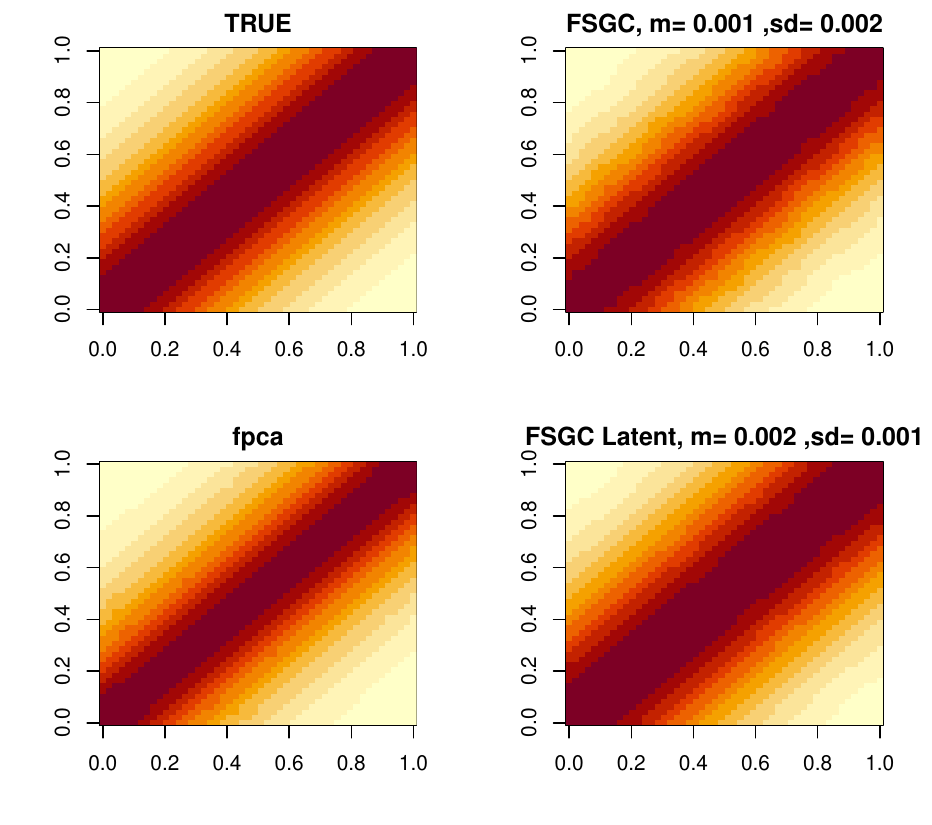}
\caption{True and Estimated average covariance surface for stationary covariance kernel, scenario C, n=1000. Average ISE (and sd) of the estimates are reported on the top of the respective images. FSGC denotes the proposed estimation method, fpca denotes FPCA on the observed curve and FSGC Latent is FPCA on latent predictions from SGCRM. }
\label{fig:fig17}
\end{figure}

\begin{figure}[H]
\centering
\includegraphics[width=1\linewidth , height=0.9\linewidth]{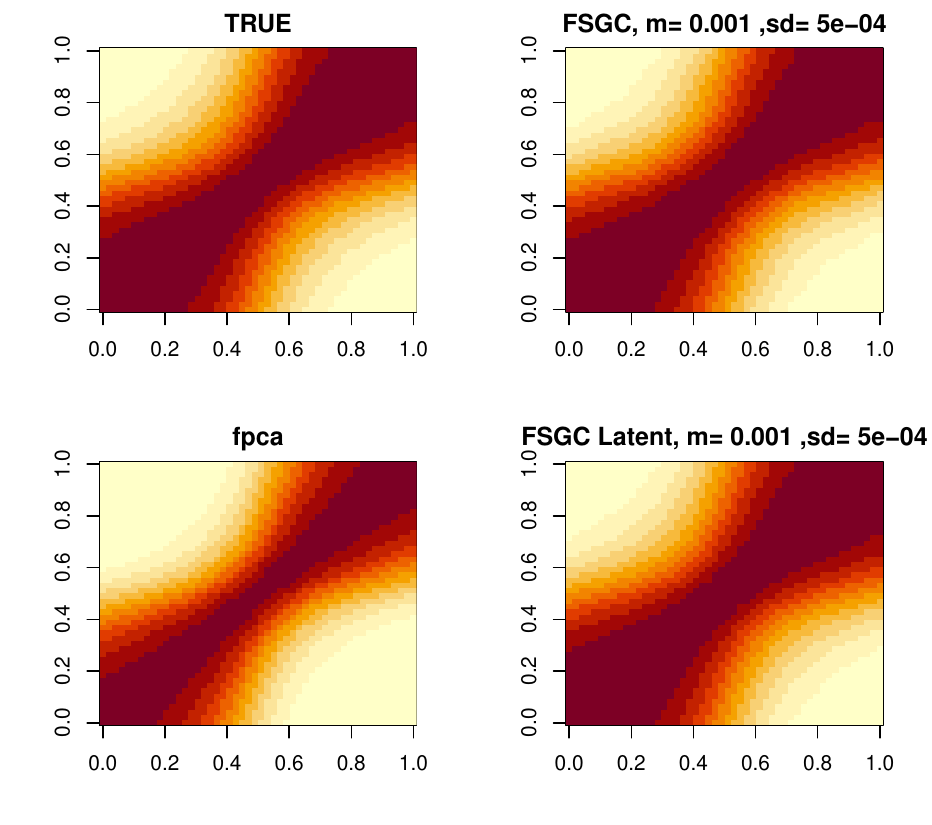}
\caption{True and Estimated average covariance surface for non-stationary covariance kernel, scenario C, n=1000. Average ISE (and sd) of the estimates are reported on the top of the respective images. FSGC denotes the proposed estimation method, fpca denotes FPCA on the observed curve and FSGC Latent is FPCA on latent predictions from SGCRM. 
}
\label{fig:fig12-tr}
\end{figure}

\subsection*{Continuous Functional Data}

\begin{figure}[H]
\centering
\includegraphics[width=1\linewidth , height=0.9\linewidth]{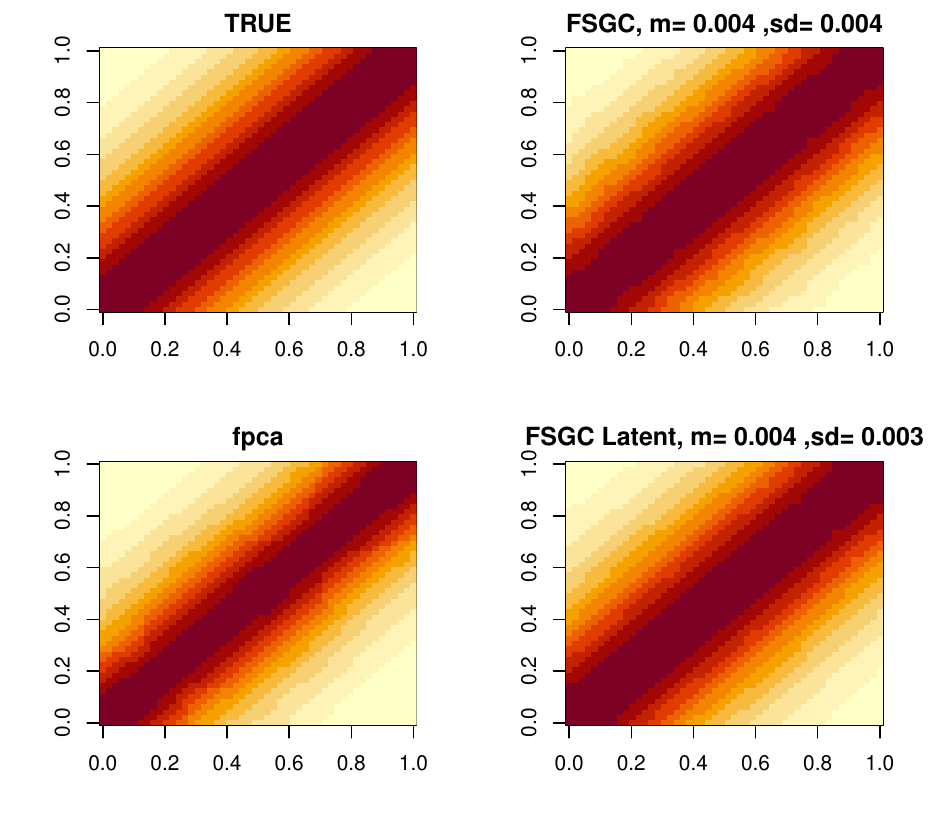}
\caption{True and Estimated average covariance surface for stationary covariance kernel, scenario D, n=100. Average ISE (and sd) of the estimates are reported on the top of the respective images. FSGC denotes the proposed estimation method, fpca denotes FPCA on the observed curve and FSGC Latent is FPCA on latent predictions from SGCRM. }
\label{fig:fig13-cont}
\end{figure}
\begin{figure}[H]
\centering
\includegraphics[width=1\linewidth , height=0.9\linewidth]{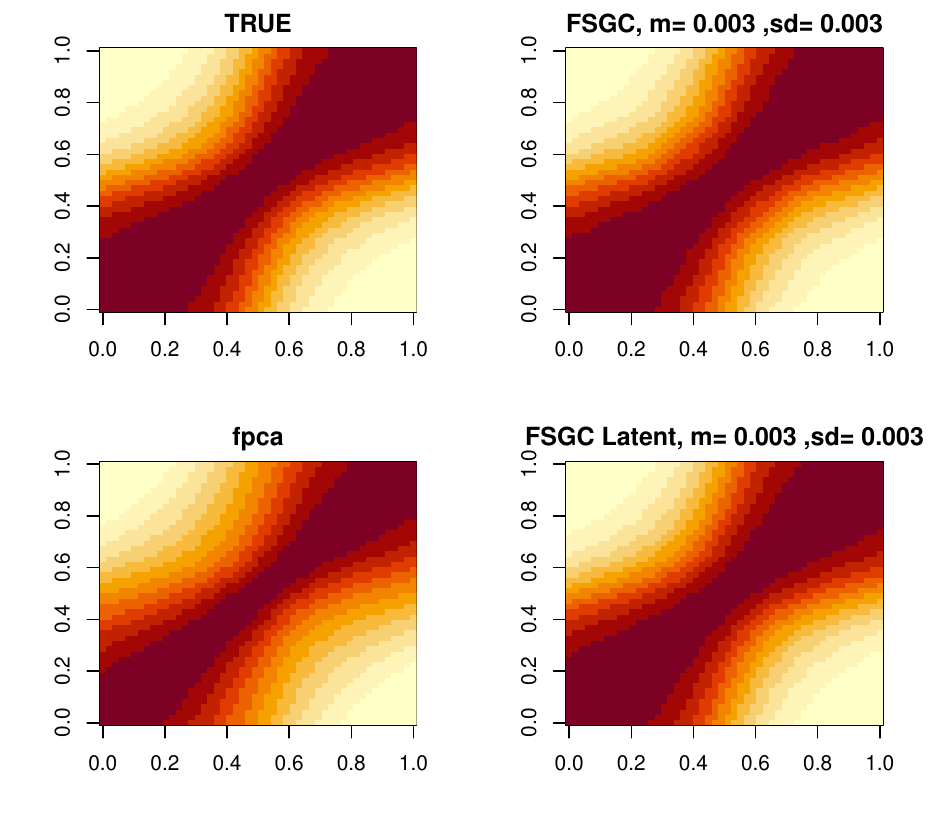}
\caption{True and Estimated average covariance surface for non-stationary covariance kernel, scenario D, n=100. Average ISE (and sd) of the estimates are reported on the top of the respective images. FSGC denotes the proposed estimation method, fpca denotes FPCA on the observed curve and FSGC Latent is FPCA on latent predictions from SGCRM. }
\label{fig:fig14}
\end{figure}

\begin{figure}[H]
\centering
\includegraphics[width=1\linewidth , height=0.9\linewidth]{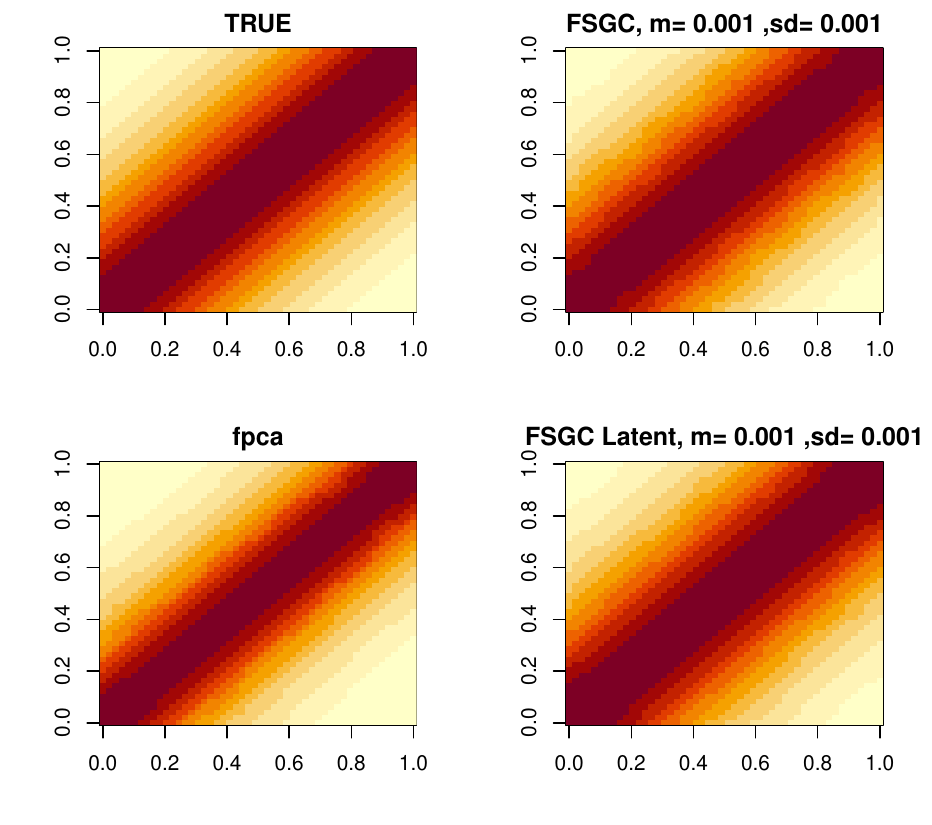}
\caption{True and Estimated average covariance surface for stationary covariance kernel, scenario D, n=500. Average ISE (and sd) of the estimates are reported on the top of the respective images. FSGC denotes the proposed estimation method, fpca denotes FPCA on the observed curve and FSGC Latent is FPCA on latent predictions from SGCRM. }
\label{fig:fig15-cont}
\end{figure}

\begin{figure}[H]
\centering
\includegraphics[width=1\linewidth , height=0.9\linewidth]{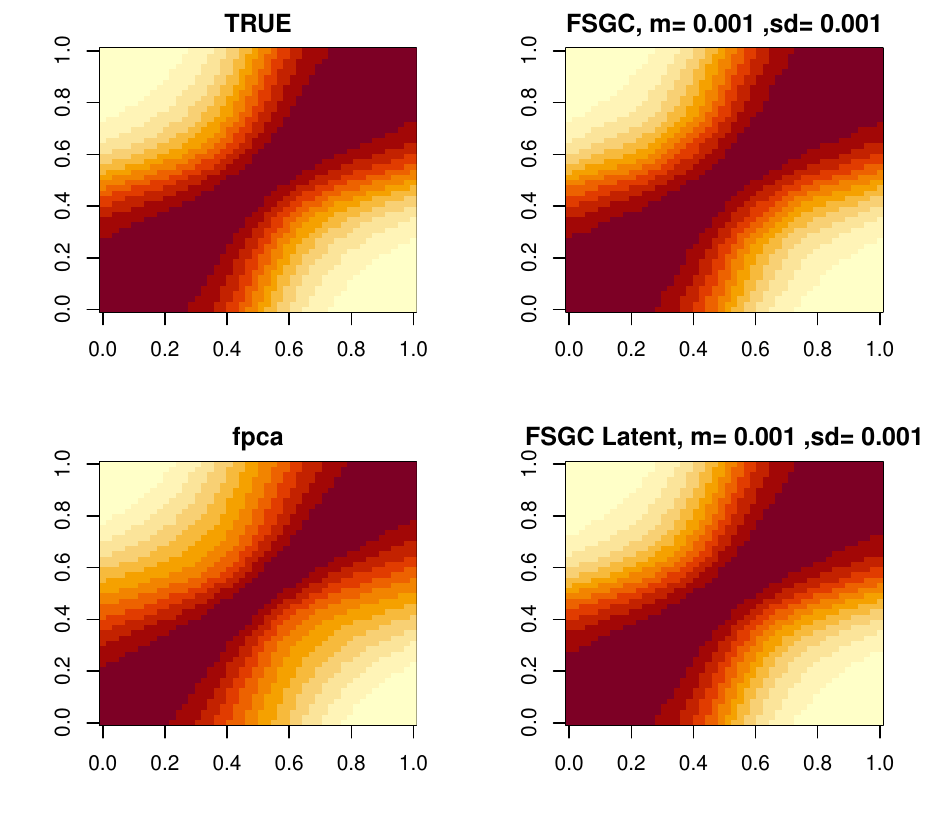}
\caption{True and Estimated average covariance surface for non-stationary covariance kernel, scenario D, n=500. Average ISE (and sd) of the estimates are reported on the top of the respective images. FSGC denotes the proposed estimation method, fpca denotes FPCA on the observed curve and FSGC Latent is FPCA on latent predictions from SGCRM.}
\label{fig:fig16}
\end{figure}

\begin{figure}[H]
\centering
\includegraphics[width=1\linewidth , height=0.9\linewidth]{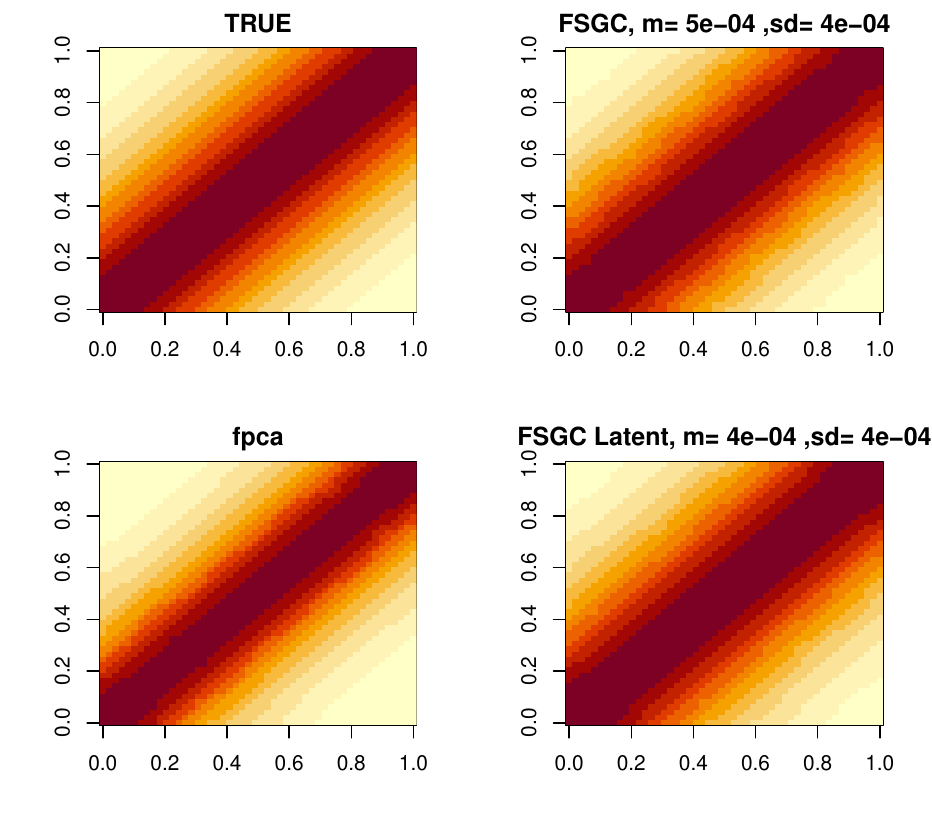}
\caption{True and Estimated average covariance surface for stationary covariance kernel, scenario D, n=1000. Average ISE (and sd) of the estimates are reported on the top of the respective images. FSGC denotes the proposed estimation method, fpca denotes FPCA on the observed curve and FSGC Latent is FPCA on latent predictions from SGCRM. }
\label{fig:fig17-cont}
\end{figure}

\begin{figure}[H]
\centering
\includegraphics[width=1\linewidth , height=0.9\linewidth]{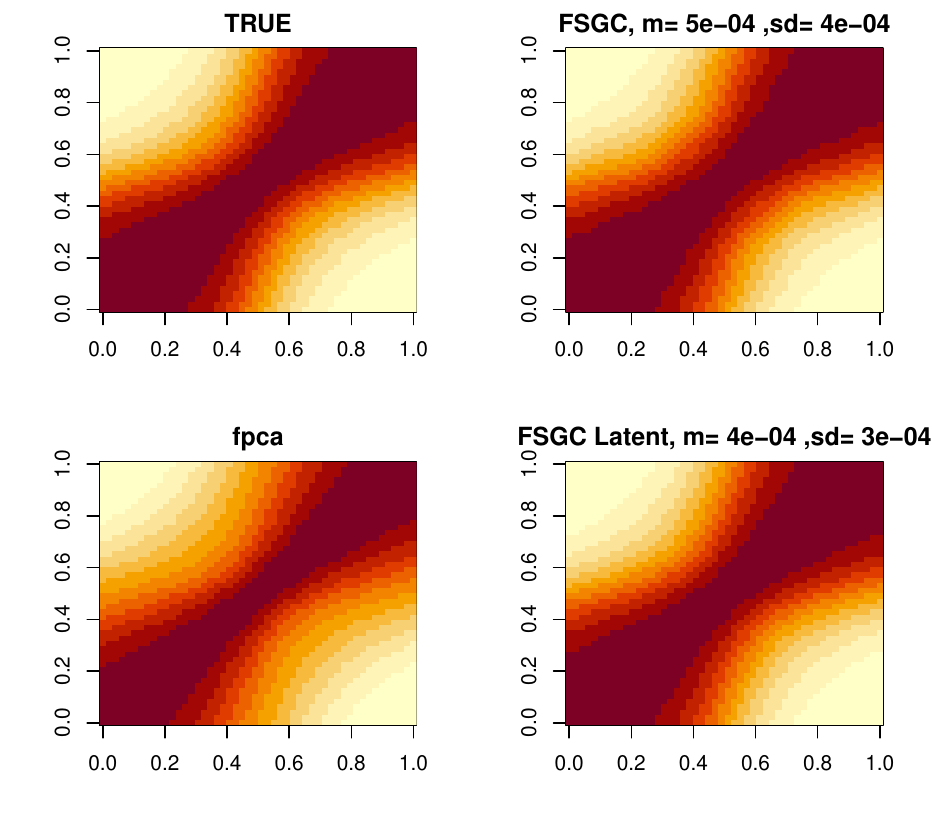}
\caption{True and Estimated average covariance surface for non-stationary covariance kernel, scenario D, n=1000. Average ISE (and sd) of the estimates are reported on the top of the respective images. FSGC denotes the proposed estimation method, fpca denotes FPCA on the observed curve and FSGC Latent is FPCA on latent predictions from SGCRM.}
\label{fig:fig12-cont}
\end{figure}


\begin{figure}[H]
\begin{center}
\begin{tabular}{ll}
\includegraphics[width=0.8\linewidth , height=0.5\linewidth]{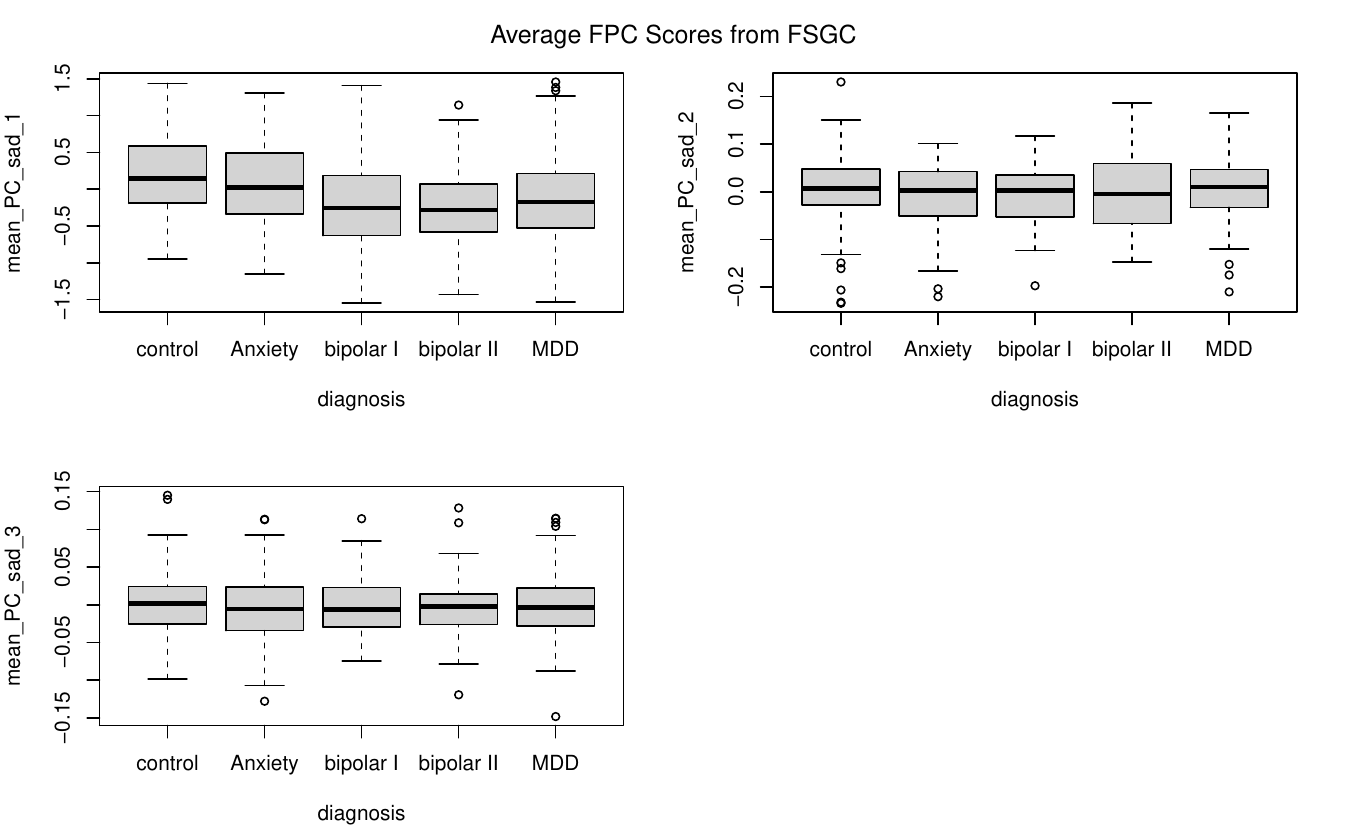} \\
\includegraphics[width=0.8\linewidth , height=0.5\linewidth]{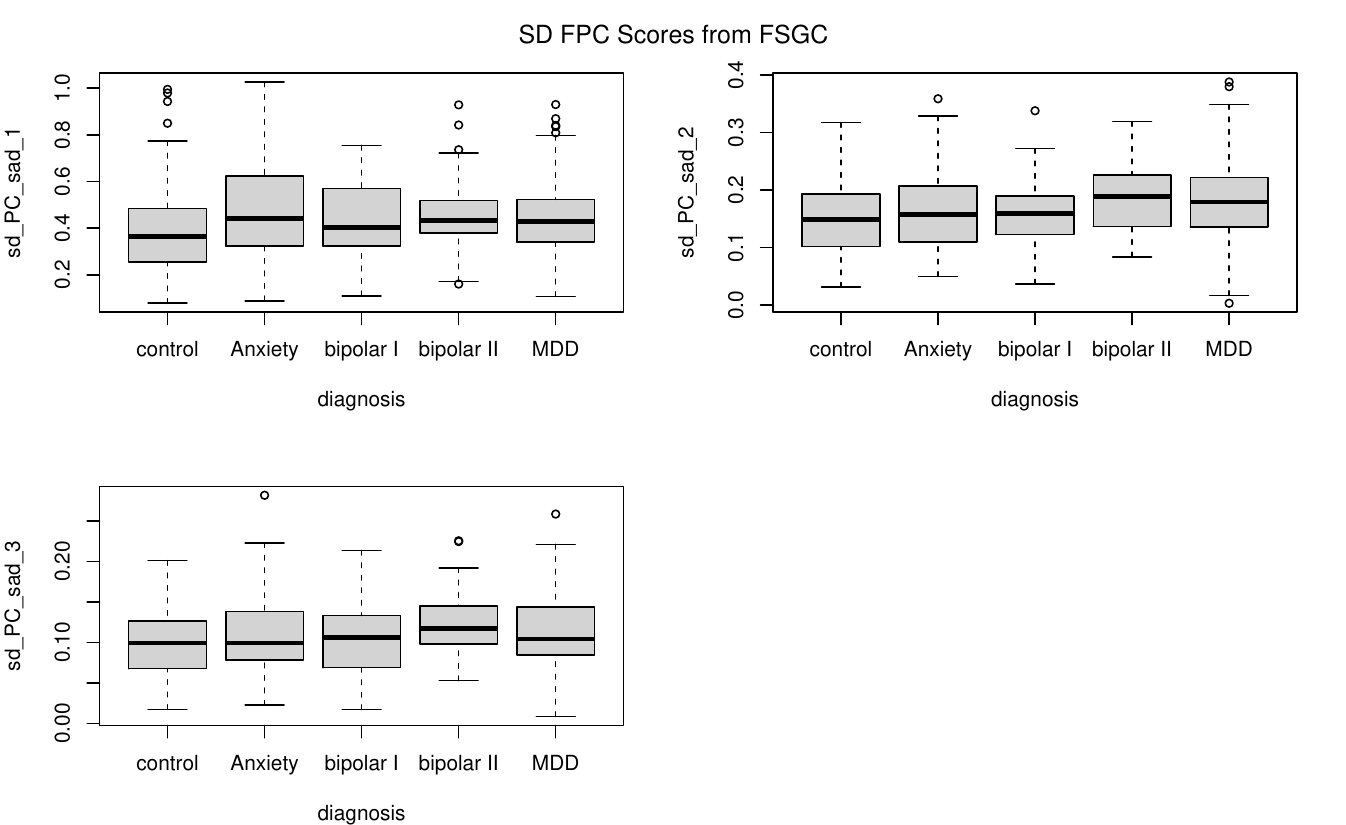}
\end{tabular}
\end{center}
\vspace{3 mm}
\caption{Distribution of mean (top panel) and SD (bottom panel) latent principal component scores of emotional states by the mood disorder groups using the proposed FSGC approach.}
\label{fig:fig4tempfinal}
\end{figure}

\begin{figure}[H]
\centering
\includegraphics[width=1\linewidth , height=0.8\linewidth]{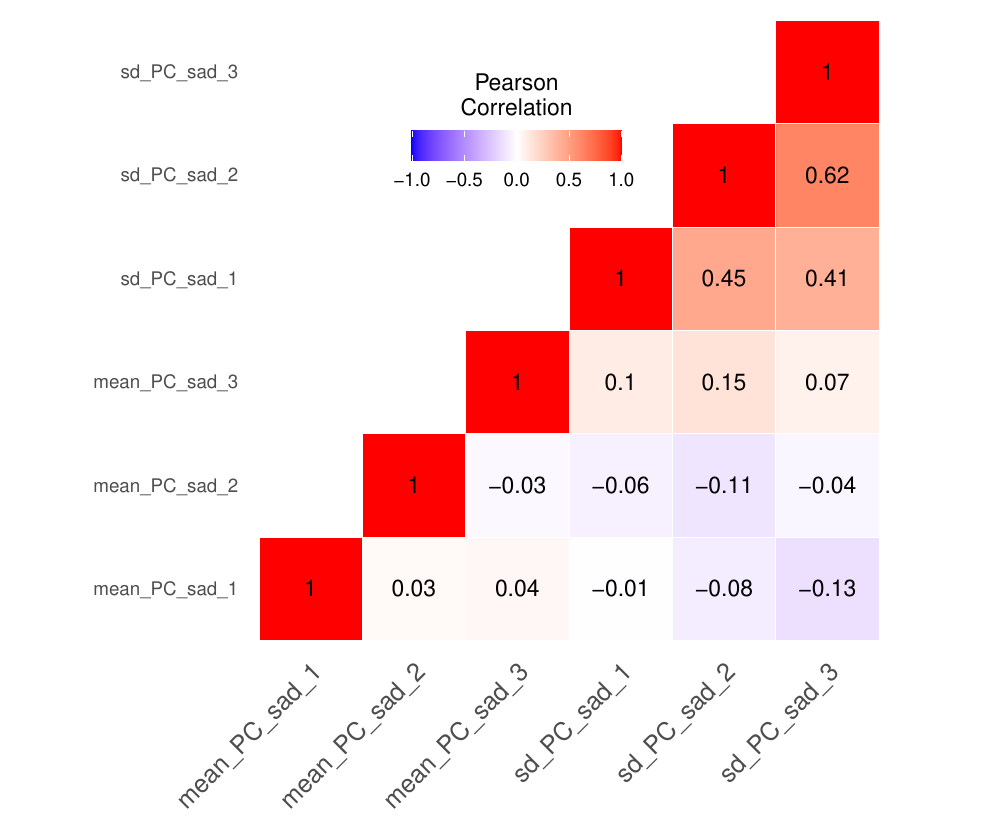}
\caption{Pearson correlation between mean and SD FPC scores of emotional states.}
\label{fig:fig4Btemp}
\end{figure}

\section{Supplementary Tables}

\begin{table}[H]
\caption{Scenario A, correlation (standard error) between estimated scores from the proposed FSGC (subscript f) and FSGC Latent (subscript l) and estimated scores from true latent curves.}
\vspace{2 mm}
\label{tab:my-table1}
\begin{tabular}{|l|l|l|l|l|l|}
\hline
Covariance Type                 & Sample size & Score1\_f     & Score1\_l     & Score2\_f     & Score2\_l     \\ \hline
\multirow{3}{*}{Stationary}     & n=100       & 0.870 (0.023) & 0.872 (0.022) & 0.800 (0.032) & 0.801 (0.032) \\ \cline{2-6} 
                                & n=500       & 0.875 (0.009) & 0.875 (0.009) & 0.806 (0.014) & 0.806 (0.014) \\ \cline{2-6} 
                                & n=1000      & 0.875 (0.007) & 0.875 (0.007) & 0.807 (0.012) & 0.807 (0.012) \\ \hline
\multirow{3}{*}{Non Stationary} & n=100       & 0.929 (0.021) & 0.929 (0.022) & 0.896 (0.028) & 0.893 (0.029) \\ \cline{2-6} 
                                & n=500       & 0.937 (0.006) & 0.937 (0.006) & 0.905 (0.009) & 0.905 (0.009) \\ \cline{2-6} 
                                & n=1000      & 0.938 (0.004) & 0.938 (0.004) & 0.906 (0.006) & 0.906 (0.007) \\ \hline
\end{tabular}

\end{table}

\begin{table}[H]
\centering
\caption{Descriptive statistics for the complete, male and female samples in the real data analysis. For continuous variable the mean and standard deviation is reported, for categorical variable the frequency in each group is mentioned.
The P-values are from two-sample t-test and Chi-Square test of association with gender.}
\vspace{3 mm}
\label{tab:my-tabler}
\centering
\small
\begin{tabular}{cccccccc}
\hline
Characteristic     & \multicolumn{1}{c|}{Complete (n=497)} & \multicolumn{1}{c|}{Male (n=195)} & \multicolumn{1}{c|}{Female (n=302)} & P value         \\ \hline
                   & Mean(sd)          & Mean(sd)     & Mean(sd)            &                 \\ \hline
Age                & 41.8 (19.5)                        &41.2 (21.7)            & 42.2(17.9)               & 0.56            \\ \hline
Diagnosis: control ($N$) & 134                       & 74           & 60                & $0.0001$          \\ 
\hline
Diagnosis: Anxiety ($N$) & 97                       & 35           & 62               &          \\ 
\hline
Diagnosis: bipolar I ($N$) & 56                       & 20           & 36                &          \\ 
\hline
Diagnosis: bipolar II ($N$) & 54                       & 22           & 32                &          \\ 
\hline
Diagnosis:  MDD  ($N$) & 156                       & 44           & 112                &          \\ 
\hline
\end{tabular}

\end{table}

\end{document}